%% file: seq_trials_arxiv_v1.tex
\documentclass[11pt,a4 paper]{article}
\usepackage{graphicx,epsf,parskip,times,amsmath,lscape,multirow,amsfonts,booktabs}
\usepackage[super,sort&compress,comma]{natbib}
\usepackage{lscape}

\newcommand\BibTeX{{\rmfamily B\kern-.05em \textsc{i\kern-.025em b}\kern-.08em
T\kern-.1667em\lower.7ex\hbox{E}\kern-.125emX}}

\usepackage{caption}
\usepackage{array}
\newcolumntype{C}[1]{>{\centering\arraybackslash}p{#1}}
\newcommand{\indep}{\perp\!\!\!\perp}

\title{Causal inference in survival analysis using longitudinal observational data: Sequential trials and marginal structural models}

\author{Ruth H. Keogh$^1$, Jon Michael Gran$^2$, Shaun R. Seaman$^3$, \\ Gwyneth Davies$^4$, Stijn Vansteelandt$^{1,5}$}

\date{{\small$^1$Department of Medical Statistics and Centre for Statistical Methodology, London School of Hygiene \& Tropical Medicine, Keppel Street, London, WC1E 7HT, UK\\
$^2$Oslo Centre for Biostatistics and Epidemiology, Department of Biostatistics, Institute of Basic Medical Sciences, University of Oslo, P.O. Box 1122 Blindern, 0317 Oslo, Norway\\
$^3$MRC Biostatistics Unit, University of Cambridge, Institute of Public Health, Forvie Site, Robinson Way, Cambridge CB2 0SR, UK\\
$^4$Population, Policy and Practice Research and Teaching Department, Great Ormond Street Institute of Child Health, University College London, London, WC1N 1EH, UK\\
$^5$Department of Applied Mathematics, Computer Science and Statistics, Ghent University, 9000 Ghent, Belgium}}

\begin{document}
	
	\maketitle

\begin{abstract}
Longitudinal observational patient data can be used to investigate the causal effects of time-varying treatments on time-to-event outcomes. Several methods have been developed for controlling for the time-dependent confounding that typically occurs. The most commonly used is inverse probability weighted estimation of marginal structural models (MSM-IPTW). An alternative, the sequential trials approach, is increasingly popular, in particular in combination with the target trial emulation framework. This approach involves creating a sequence of `trials' from new time origins, restricting to individuals as yet untreated and meeting other eligibility criteria, and comparing treatment initiators and non-initiators. Individuals are censored when they deviate from their treatment status at the start of each `trial' (initiator/non-initiator) and this is addressed using inverse probability of censoring weights. The analysis is based on data combined across trials. We show that the sequential trials approach can estimate the parameter of a particular MSM, and compare it to a MSM-IPTW with respect to the estimands being identified, the assumptions needed and how data are used differently. We show how both approaches can estimate the same marginal risk differences. The two approaches are compared using a simulation study. The sequential trials approach, which tends to involve less extreme weights than MSM-IPTW, results in greater efficiency for estimating the marginal risk difference at most follow-up times, but this can, in certain scenarios, be reversed at late time points. We apply the methods to longitudinal observational data from the UK Cystic Fibrosis Registry to estimate the effect of dornase alfa on survival.
\end{abstract}

\section{Introduction}

Longitudinal observational data on patients can allow for the estimation of treatment effects over long follow-up periods and in diverse populations. A key task when aiming to estimate causal effects from observational data is to account for confounding of the treatment-outcome association. With longitudinal data on treatment and outcome, the treatment-outcome association may be subject to so-called time-dependent confounding. Such confounding is present when there are time-dependent covariates affected by past treatment, also affecting later treatment use and outcome. Over the last two decades a large statistical literature has built up on methods for estimating causal treatment effects in the presence of time-dependent confounding, and their use in practice is becoming more widespread.

In this paper we focus on estimating the effects of a longitudinal treatment regime on a time-to-event outcome, using longitudinal observational data on treatment and covariates obtained at approximately regular visits, alongside the time-to-event information. When there is time-dependent confounding, standard methods for survival analysis, such as Cox regression with adjustment for baseline or time-updated covariates, do not in general enable estimation of causal effects. Several methods have been described for estimating the causal effects of longitudinal treatment regimes on time-to-event outcomes. Marginal structural models (MSMs) estimated using inverse probability of treatment weighting (IPTW) were introduced by Robins et al.\cite{Robins:2000} and extended to survival outcomes by Hernan et al.\cite{Hernan:2000} through marginal structural Cox models. A review by Clare et al. in 2019\cite{Clare:2019} found this approach to be by far the most commonly used in practice, and we refer to this as the MSM-IPTW approach. An alternative but related approach, which we refer to as the `sequential trials' approach, was described by Hernan et al. in 2008\cite{Hernan:2008} and then Gran et al. in 2010\cite{Gran:2010}.

In the sequential trials approach, artificial `trials' are mimicked from a sequence of new time origins which could, for example, be defined by study visits at which new information is recorded for each individual who remains under observation. At each time origin individuals are divided into those who have just initiated the treatment under investigation and those who have not yet initiated the treatment. Within each mimicked trial, individuals are artificially censored at the time at which their treatment status deviates from what it was at the time origin, if such deviation occurs. Inverse probability of censoring weighting is used to account for dependence of this artificial censoring on time-dependent characteristics. Effects of being treated or untreated starting from each time origin can then be estimated using, for example, weighted pooled logistic or Cox regression. There have been several applications of the sequential trials approach. Danaei et al.(2013) \cite{Danaei:2013SMMR} applied it to electronic health records data to estimate the effect of statins on occurrence of coronary heart disease. Clark et al. (2015)\cite{Clark:2015} studied whether onset of impaired sleep is followed by changes in health related behavior. Bhupathiraju et al. (2017)\cite{Bhupathiraju:2017} investigated the effect of hormone therapy use on chronic disease risk in the Nurses' Health Study. Some papers have also used the sequential trials approach alongside other methods and compared the results empirically. Suttorp et al (2015)\cite{Suttorp:2015} studied the effect of high dose of a particular agent used in kidney disease patients on all-cause mortality with both a sequential trials approach and MSM-IPTW using data from a longitudinal cohort. Thomas et al. (2020)\cite{Thomas:2020} used the sequential trials approach together with sequential stratification and time-dependent propensity score matching when studying the effect of statins on cardiovascular outcomes using the Framingham Offspring cohort.

The sequential trials approach is attractive because it mimics a series of randomized trials in a fairly straightforward and explicit manner. Several recent papers have advocated to  emulate a `target trial' in investigations of causal effects using observational data \citep{Hernan:2016AJE,Hernan:2016JCE,Garcia:2017, Hernan:2018,Didelez:2016}. The target trial is the hypothetical randomized controlled trial that one would like to conduct. The use of target trials provides a structured framework that guides the steps of an investigation, with the target trial protocol including specification of eligibility criteria, time origin, and the estimands of interest. There is strong link between the target trial framework of Hernan and Robins (2016)\cite{Hernan:2016AJE} and the general roadmap for causal inference proposed by Petersen and van der Laan (2014)\cite{Petersen:2014}. These approaches help to avoid biases in estimation of treatment effects using observational data, and to clarify assumptions and methodology.

Applications of the sequential trials approach, including in the original papers of Hernan et al.\cite{Hernan:2008} and Gran et al.\cite{Gran:2010}, have focused on estimation of hazard ratios. However, it has not been discussed in detail what the formal causal interpretation of this hazard ratio is and Gran and Aalen (2019)\cite{Gran:2019} highlighted the need for further work to establish the causal estimand being estimated. This is important both to clarify the interpretation and to inform simulation studies to study the properties of the estimators. Karim et al. (2018)\cite{Karim:2018} compared the sequential trials approach with MSM-IPTW based on Cox models using simulations, but did not account for the fact that the two analyses estimate different quantities, which are therefore not directly comparable \citep{Gran:2019}. The aim of this paper is to formally describe what and how causal estimands can be identified using the sequential trials approach in a time-to-event setting and to contrast these with estimands typically identified using MSM-IPTW. We show that the causal parameter estimated from the sequential trials approach can be described in terms of the parameters of a particular marginal structural model. We also establish more broadly what estimands the two approaches are suitable for identifying, what assumptions are needed and how the data are used differently. 

In the causal inference literature on the sequential trials approach and MSM-IPTW for time-to-event outcomes, the focus has mostly been on Cox proportional hazard models and estimation of hazard ratios. However, Aalen additive hazard models for hazard differences are increasingly being used, both in general to avoid the proportional hazards assumption and in causal inference due to their useful property of being collapsible \citep{Martinussen:2013}. We will therefore outline methods in terms of both proportional and additive hazards models. As hazard ratios, and generally also hazard differences, have been shown not to have a straightforward causal interpretation \citep{Hernan:2010,Aalen:2015,Martinussen:2020}, we demonstrate how marginal survival probabilities can be obtained under both hazard models for both the sequential trials and MSM-IPTW approaches, using a simple standardization procedure. This enables results from hazard models to be converted into more meaningful causal quantities such as risk differences or risk ratios. The two approaches are compared in a simulation study, in which we assess their relative efficiency for estimation of the same quantities, making use of the properties of additive hazards models \citep{Keogh:2021}.

The paper is organised as follows. In Section \ref{sec:motivation} we describe a motivating example, in which the aim is to estimate the effect of long term use of the treatment dornase alfa on the composite outcome of death or transplant for people with cystic fibrosis (CF), using longitudinal data from the UK Cystic Fibrosis Registry. In Section \ref{sec:notationandtt} we introduce notation and describe the components of a general target trial in this context. In Section \ref{sec:ttemulation} we outline estimation using the MSM-IPTW approach and the sequential trials approach, showing how the latter can also be understood as fitting a particular MSM, and in Section \ref{sec:comparison} we discuss in detail the similarities and differences between the two approaches. The two approaches are then compared using a simulation study in Section \ref{sec:sim}, and applied to the motivating example in Section \ref{sec:example}. Accompanying R code for performing the simulation is provided at https://github.com/ruthkeogh/sequential\_trials. We conclude with a discussion in Section \ref{sec:discussion}. Our focus is on MSM-IPTW and the sequential trials approach. Petersen et al. (2007)\cite{Petersen:2007a} described history-adjusted MSMs (HA-MSM), which are connected to both approaches, but which have not been used much in practice, perhaps due to initial criticisms of the method \citep{Robins:2007,Petersen:2007b}. In Section \ref{sec:discussion} we discuss the link between HA-MSM and MSM-IPTW and the sequential trials approach. 

\section{Motivating example}
\label{sec:motivation}

As a motivating example we will investigate the long term impacts of a treatment used in CF on the composite time-to-event outcome of death or transplant. CF is an inherited, chronic, progressive condition affecting around 10,500 individuals in the UK and over 70,000 worldwide \citep{CFT:2019,Rowe:2005,MacNeill:2016}. Survival in CF has improved considerably over recent decades and the estimated median survival age for a person born with CF in the UK today is 51.6 for males and 45.7 for females \citep{Dodge:2007,Keogh:2018JCF2,CFT:2019}.

One of the most common symptoms of CF is a build up of mucus in the lungs, which leads to an increased prevalence of bacterial growth in the airways and a decline in lung function \citep{Pressler:2008}. Therefore, it is common for people with CF to routinely use aerosolized mucoactive agents, which help to break down the layer of mucus in the lungs making clearance easier. Recombinant human deoxyribonuclease, commonly known as dornase alfa (DNase), is one such mucolytic treatment which was authorised for use in January 1994. DNase is the most commonly used treatment for the pulmonary consequences of CF in the UK, used by 67.6\% of people in 2019 \citep{CFT:2019}. It is a nebulised treatment, administered daily on a long term basis. The efficacy of DNase for health outcomes in CF has been studied in several randomized controlled trials \citep{Yang:2018}. Most trials had short term follow-up and focused on the impact of DNase on lung function. Seven trials from the 1990s investigated mortality as a secondary outcome but follow-up was short and findings inconclusive; a meta-analysis based on these studies gave a risk ratio estimate of 1.70 (95\% CI 0.70 to 4.14) comparing users versus non-users \citep{Yang:2018}. 

The effect of DNase use on lung function and requirement for intravenous antibiotics was investigated using observational data from the UK CF Registry \citep{Newsome:2018SIM,Newsome:2019JCF}. We now use the UK CF Registry to investigate the impact of DNase on the composite outcome of death or transplant through emulation of a target trial. The majority of transplants in people with CF are lung transplants, however we consider any transplant here. The UK CF Registry is a national, secure database sponsored and managed by the Cystic Fibrosis Trust \citep{Taylor-Robinson:2017}. It was established in 1995 and records demographic data and longitudinal health data on nearly all people with CF in the UK, to date capturing data on over 12,000 individuals. Data are collected in a standardized way at designated (approximately) annual visits on over 250 variables in several domains, and have been recorded using a centralised database since 2007. In this study we use data from 2008 to 2018, plus some data on prior years to define eligibility according to our criteria outlined below. At each annual visit it is recorded whether or not an individual had been using DNase in the past year. We identified potential confounders of the association between DNase use and death or transplant based on expert clinical input, and these include both time-fixed and time-dependent variables - see Section \ref{sec:example}. We focus on the impact of initiating and continuing use of DNase on the risk of death or transplant up to 11 years of follow-up compared with not using DNase. In Section \ref{sec:example} we describe the target trial protocol for addressing this question and how it is emulated using the observational data. 

\section{Target trial and notation}
\label{sec:notationandtt}

We begin by outlining a general target trial protocol in the context of estimating the effect of a longitudinal treatment regime on a time-to-event outcome, using the elements recommended by Hernan and Robins\cite{Hernan:2016AJE}. This is a protocol for a hypothetical trial that, if possible, could have been performed to estimate the estimand of interest. We will later focus on how to emulate this trial when we have longitudinal observational data on treatment, covariates and the outcome.

\emph{Eligibility criteria}

Individuals are eligible to enter the trial from the first time that they meet a specified criteria (for example based on age, clinical characteristics and counter-indicators to the treatment in question) during a particular recruitment period. Individuals who have previously used the treatment or treatments under study are not eligible. 

\emph{Treatment strategies}

We consider a binary treatment $A$ (treatment or control), and let $\underline{A}_0$ denote treatment status from the time they enter the trial (time 0) onwards. Individuals are assigned to the treatment or control group, which they are asked to sustain throughout follow-up. Those who sustain the treatment throughout follow-up have $\underline{A}_0=1$, and those who do not use treatment throughout follow-up have $\underline{A}_0=0$.

\emph{Assignment procedures}

At the time of entering the trial, individuals are randomized to the treatment group ($\underline{A}_0=1$) or the control group ($\underline{A}_0=0$). In a real trial, double- or single-blind assignment of the treatment would be desirable in contexts where blinding is feasible. However, it is not realistic to emulate blinding when the target trial is emulated using the observational data at hand. Therefore in the target trial both participants and their clinical teams are aware of their treatment group.  

\emph{Follow-up period}

Individuals are followed up from randomization, at $t=0$, until the time of the event of interest $T$, the time horizon of interest $\tau$, or censoring $C$, whichever occurs first. 

\emph{Outcome}

The observed outcome for a given individual is $\tilde T=\mathrm{min}( T,C, \tau)$. 

\emph{Causal estimands of interest}

Let us define $T^{\underline{a}_0=1}$ as the counterfactual event time had an individual received the treatment from time zero onwards (`always treated') and $T^{\underline{a}_0=0}$ as the counterfactual event time had an individual not received the treatment from time zero onwards (at least for the follow-up time of interest) (`never treated'). We can then define our causal estimands of interest as contrasts between functions of the distributions of $T^{\underline{a}_0=1}$ and $T^{\underline{a}_0=0}$. These correspond to per-protocol effects because they involve comparisons between outcomes if treatment level $a$ ($a=0,1$) were sustained throughout the follow-up period. 

Effects of treatments on time-to-event outcomes can be quantified in a number of ways. A common measure is the ratio of hazards between the treatment and control groups, made under the assumption that hazards are proportional over the period of follow-up, $h_{T^{\underline{a}_0=1}}(t)/h_{T^{\underline{a}_0=0}}(t)=e^{\beta}$, where $h_{T^{\underline{a}_0=a}}(t)$ denotes the hazard in the hypothetical world in which all individuals are observed in treatment group $a$ ($a=0,1$). However, as hazard ratios do not have a straightforward causal interpretation, it is additionally advisable to consider causal contrasts between risks \citep{Hernan:2010,Aalen:2015,Martinussen:2020}. In this paper we will focus on the risk difference as the primary estimand of interest:
\begin{equation}
    \Pr(T^{\underline{a}_0=1}>\tau)-\Pr(T^{\underline{a}_0=0}>\tau).
    \label{eq:estimand}
\end{equation}
This risk difference contrasts the risk of the event by the time horizon $\tau$, under the conditions of being `always treated' versus `never treated'. There may be several time horizons of interest, i.e. different values of $\tau$, and we let $\tau_{\mathrm{max}}$ denote the maximum time horizon for which the risk difference is to be obtained. Alternative quantities with a causal interpretation are the risk ratio $\Pr(T^{\underline{a}_0=1}>\tau)/\Pr(T^{\underline{a}_0=0}>\tau)$ and contrasts between restricted mean survival times $E(\mathrm{min}(T^{\underline{a}_0},\tau))$ \citep{Royston:2013}.

\emph{Analysis plan for the target trial}

If the target trial in fact could have been implemented, the estimation of the causal risk difference in (\ref{eq:estimand}) would make use of analysis methods for randomized controlled trials. In a randomized controlled trial in the absence of non-adherence or informative drop-out, the counterfactual event times are independent of the treatment, i.e. $T^{\underline{a}_0} \indep A_0$ for all $a_0$. Under this assumption, the causal risk difference of interest is equal to the observed risk difference:
\begin{equation}
    \Pr(T^{\underline{a}_0=1}>\tau)-\Pr(T^{\underline{a}_0=0}>\tau)=\Pr(T>\tau|A_0=1)-\Pr(T>\tau|A_0=0).
\end{equation}
The risks $\Pr(T>\tau|A_0=a)$ ($a=0,1$) could, for example, be estimated nonparametrically using the Kaplan-Meier estimator. In practice, non-adherence and informative drop-out are often present, and the analysis should take account of this, with IPTW being one possible approach.

\section{Emulating the target trial using longitudinal observational data}
\label{sec:ttemulation}

\subsection{Observational data set-up}

Let us now assume a setting in which individuals are observed at regular `visits' (data collection times) over a particular calendar period, reflecting the type of data that arise in our motivating example and in similar data sources. Let $t=0$ denote the start of follow-up, defined as the time at which a patient meets the eligibility criteria. We assume a `closed cohort' such that all individuals in the study cohort were included at the same calendar time. In practice, some historical data are typically required to establish whether an individual meets the eligibility criteria at time $t=0$, e.g. past treatment status. The cohort is assumed to be followed up to (at least) the maximum time horizon of interest for the risk difference, $\tau_{\mathrm{max}}$. Information on treatment status and other covariates is observed at visits $k=0,1,\ldots,K$, and the time of visit $k$ is denoted $t_k$. Without loss of generality, we assume that $t_k=k$, for $k \geq 0$, and $K=\lfloor \tau_{\mathrm{max}}\rfloor$denotes the time of most recent visit before time $\tau_{\mathrm{max}}$ ($\lfloor \tau_{\mathrm{max}}\rfloor<\tau_{\mathrm{max}}$). At each visit $k$ we observe a binary treatment status $A_{k}$ and a set of time-dependent covariates $L_{k}$. To simplify the notation, each $L_k$ ($k=0,1,\ldots$) also includes any time-fixed covariates. A bar over a time-dependent variable indicates the history, that is $\bar{A}_{k}=\{A_0,A_1,\ldots,A_k\}$ and $\bar{L}_{k}=\{L_0,L_1,\ldots,L_k\}$. An underline indicates the future, so that $\underline{A}_{k}=\{A_k,A_{k+1},\ldots\}$ denotes treatment status from time $k$ onwards. 

\begin{figure}
	\centering
		\includegraphics[scale=0.8]{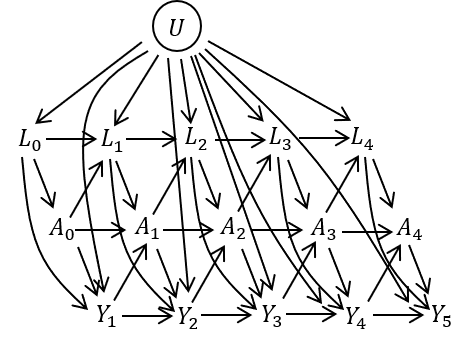}
			\caption{Directed acyclic graph (DAG) illustrating relationships between treatment $A$, time-dependent covariates $L$, discrete time outcome $Y$, and unmeasured covariates $U$. Baseline covariates $Z$ are omitted from the diagram but are assumed to potentially affect all other variables.}\label{fig:dag}
\end{figure}

The data set-up is illustrated in the directed acyclic graph (DAG) in Figure \ref{fig:dag}, using a discrete-time setting where $Y_k=I(t_{k-1} \leq T< t_{k})$ is an indicator of whether the event occurs between visits $k-1$ and $k$. One can imagine extending the DAG by adding a series of small time intervals between each visit, at which events are observed (but not $A$ or $L$, which are assumed constant between visits). As the time intervals become very small we approach the continuous time setting. From the DAG, we can see that $L_k$ are time-dependent confounders. Time-dependent confounding occurs when there are time-dependent covariates that predict subsequent treatment use, are affected by earlier treatment, and affect the outcome through pathways that are not just through subsequent treatment. The DAG also includes a variable $U$, which has direct effects on $L_k$ and $Y_k$ but not on $A_k$. $U$ is an unmeasured individual frailty and we include it because it is realistic that such individual frailty effects exist in practice. Because $U$ is not a confounder of the association between $A_k$ and $Y_k$ (after controlling for the observed confounders), the fact that it is unmeasured does not affect our ability to estimate causal effects of treatments. The DAG could be extended in various ways, for example, we could incorporate long term effects of $L$ on $A$ and vice versa by adding arrows from $L_{k}$ to $A_{k+1}$ and from $A_k$ to $L_{k+2}$. Long term effects of $A$ and $L$ on survival could also be added, for example by adding arrows from $L_{k}$ and $A_{k}$ to $Y_{k+2}$.

\subsection{The emulated trial}

Next, we outline how the target trial outlined in Section \ref{sec:notationandtt} is emulated using the observational data described above. Each element of the target trial protocol should be considered. The eligibility criteria, treatment strategies, assignment strategies, and outcome in the emulated trial should be the same as far as possible as those in the target trial. An iterative process is appropriate when designing the target trial protocol, to take into consideration knowledge about the observational data. In some settings there may be some differences in eligibility criteria in the target trial and the emulated trial - for example, if the study pertains to individuals diagnosed with a particular infection, in the target trial a laboratory-confirmed diagnosis may be specified, whereas this may not possible in the emulated trial if diagnosis route is not recorded in the available observational data. There may also be differences in some aspects of treatment, for example the target trial may specify the dose of the treatment, whereas dose information may not be available in the observational data and so the emulated trial is pragmatic about dose. Making clear the differences between the target trial and the emulated trial can help us to understand possible sources of bias when we emulate the target trial using observational data. 

The causal estimand of interest for the emulated trial is the same as that in the target trial. In our setting, the causal estimand compares marginal risks in a follow-up period $\tau$ under two treatment regimes: always treated ($\underline{a}_0=1$) or never treated ($\underline{a}_0=0$). In the observational data, individuals who are untreated initially can subsequently start treatment and vice versa. Treatment initiation and treatment switching may depend on time-dependent patient characteristics that also affect the outcome, as illustrated in the DAG in Figure \ref{fig:dag}. This must be addressed in the analysis, and the analysis stage is where we find the biggest differences between the emulated trial and target trial. However, we emphasise that the causal estimand in the emulated trial is the same as in the target trial. The time-dependent confounding is addressed in different ways in the MSM-IPTW approach and the sequential trials approach. We now describe how to estimate the causal risk difference (expression (\ref{eq:estimand})) using these two approaches. 

\subsection{Analysis using MSM-IPTW}
\label{sec:msm}

In the MSM-IPTW approach, a MSM is specified for the hazard. The counterfactual hazard under the possibly counter-to-fact treatment regime $\underline{a}_{0}$ is denoted $h_{T^{\underline{a}_{0}}}(t)$. In a slight abuse of standard notation we let $\lfloor t \rfloor$ denote the time of the most recent visit \emph{before} time $t$ (so $\lfloor t \rfloor$ is always $<t$), and $\bar{a}_{\lfloor t \rfloor}$ denotes treatment pattern up to the most recent visit prior to $t$. The MSM is usually assumed to take the form of a Cox proportional hazards model \citep{Cox:1972}
\begin{equation}
h_{T^{\underline{a}_{0}}}(t)=h_{0}(t)e^{g(\bar{a}_{\lfloor t \rfloor};\beta_A)},
\label{eq:msm.cox}
\end{equation}
where $h_{0}(t)$ is the baseline counterfactual hazard, and $g(\bar{a}_{\lfloor t \rfloor};\beta_A)$ is a function of treatment pattern $\bar{a}_{\lfloor t \rfloor}$ to be specified, and $\beta_A$ is a vector of log hazard ratios. However, the hazards could take various other forms and we will also consider MSMs based on Aalen's additive hazard model \citep{Aalen:1989,Aalen:2008}, which has the form
\begin{equation}
h_{T^{\underline{a}_{0}}}(t)=\alpha_{0}(t)+g(\bar{a}_{\lfloor t \rfloor};\alpha_A(t)),
\label{eq:msm.aalen}
\end{equation}
where $\alpha_{0}(t)$ is the baseline counterfactual hazard and $\alpha_A(t)$ is a vector of time-dependent coefficients. In both hazard models the $g(\cdot)$ function equals zero when $\bar{a}_{\lfloor t \rfloor}=0$. In a simple version of the MSM, the hazard at time $t$ is assumed to depend only on the current level of treatment: $g(\bar{a}_{\lfloor t \rfloor};\beta_A)=\beta_A a_{\lfloor t \rfloor}$ or $g(\bar{a}_{\lfloor t \rfloor};\alpha_A(t))=\alpha_A(t) a_{\lfloor t \rfloor}$. Other options include that the hazard depends on duration of treatment, $g(\bar{a}_{\lfloor t \rfloor};\beta_A)=\beta_A \sum_{k=0}^{\lfloor t \rfloor}a_{k}$ or $g(\bar{a}_{\lfloor t \rfloor};\alpha_A(t))=\alpha_A(t) \sum_{k=0}^{\lfloor t \rfloor}a_{k}$; or on the history of treatment in a more general way, $g(\bar{a}_{\lfloor t \rfloor};\beta_A)=\sum_{k=0}^{\lfloor t \rfloor}\beta_{Ak} a_{k}$ or $g(\bar{a}_{\lfloor t \rfloor};\alpha_A(t))=\sum_{k=0}^{\lfloor t \rfloor}\alpha_{Ak}(t) a_{k}$. 

The MSM cannot be estimated directly from the observed data because of time-dependent confounding by $L_k$. In the MSM-IPTW approach, individuals are assigned time-dependent weights and the MSM is fitted to the observed weighted data. The weight at time $t$ for a given individual is the inverse of the probability of their observed treatment pattern up to time $t$ given their time-dependent covariate history, $\bar{L}_{\lfloor t \rfloor}$. These weights can be large for some individuals, which induces large uncertainty in estimates from the MSM, and stabilized weights are typically used instead. The use of MSMs estimated using IPTW to estimate causal effects of joint treatments over time involves the four key assumptions of no interference, positivity, consistency, and conditional exchangeability (i.e. no unmeasured confounding) \citep{Robins:2000,Vanderweele:2009,Daniel:2013}. Details on the weights and the assumptions are provided in the Supplementary Material (Sections A1 and A2). Weights are also discussed in detail in the context of the simulation study in Section \ref{sec:sim}.

The MSM can also be extended to include conditioning on baseline covariates $L_0$:
\begin{equation}
h_{T^{\underline{a}_{0}}}(t| L_0)=h_{0}(t)e^{g(\bar{a}_{\lfloor t \rfloor};\beta_A)+\beta_L  L_0}
\label{eq:msm.cox.L}
\end{equation}
or
\begin{equation}
h_{T^{\underline{a}_{0}}}(t|L_0)=\alpha_{0}(t)+g(\bar{a}_{\lfloor t \rfloor};\alpha_A(s))+\alpha_L(t) L_0.
\label{eq:msm.aalen.L}
\end{equation}
When covariates $ L_0$ are included in the MSM, they can also be included in the model in the numerator of the stabilized weights. 

The survival probabilities required for the estimand in (\ref{eq:estimand}) can be estimated using the relation between the hazard and survivor functions. For the Cox MSM in (\ref{eq:msm.cox}) we have
	\begin{equation}
	\begin{split}
	\Pr(T^{\underline{a}_{0}}>\tau)&=\exp\left\{-e^{g(a_{0};\tilde\beta_A)}\int_{0}^{1}h_{0}(s)ds-e^{ g(\bar{a}_{1};\tilde\beta_A)}\int_{1}^{2}h_{0}(s)ds\cdots -e^{g(\bar{a}_{\lfloor t \rfloor};\tilde\beta_A)}\int_{\lfloor \tau \rfloor}^{\tau}h_{0}(s)ds\right\},
	\end{split}
	\label{eq:surv.cox.msm}
	\end{equation}
	where the integrated (cumulative) baseline hazards can be estimated using an IPTW Breslow's estimator. Based on the Aalen MSM in (\ref{eq:msm.aalen}) we have
	\begin{equation}
		\Pr(T^{\underline{a}_{0}}>\tau)=\exp\left(-\int_{0}^{t}\alpha_{0}(s)ds-\int_{0}^{1}g(a_{0};\alpha_A(s))ds-\int_{1}^{2}g(\bar{a}_{1};\alpha_A(s))ds\cdots -\int_{\lfloor \tau \rfloor}^{\tau}g(\bar{a}_{\lfloor \tau \rfloor};\alpha_A(s))ds\right),
		\label{eq:surv.aalen.msm}
	\end{equation}
where the integrals are obtained as the cumulative coefficients estimated in the Aalen additive hazard model fitting process. The probabilities of interest under the longitudinal treatment regimes of `always treated' and `never treated' ($\Pr(T^{\underline{a}_{0}=a}>\tau)$, $a=0,1$) are the obtained by setting $\underline{a}_{0}=a$ in (\ref{eq:surv.cox.msm}) or (\ref{eq:surv.aalen.msm}). These probabilities are marginal and they refer to the population of $n$ individuals meeting the target trial eligibility criteria at $t=0$; we denote this population by $\mathcal{C}_0$. The MSMs including baseline covariates, as in (\ref{eq:msm.cox.L}) and (\ref{eq:msm.aalen.L}), can be used to obtain the conditional probabilities $\Pr(T^{\underline{a}_{0}=a}\geq \tau|L_{0})$ for each individual in $\mathcal{C}_0$. Estimates of the marginal survival probabilities $\Pr(T^{\underline{a}_{0}}>\tau)$ can then be obtained using empirical standardization:
\begin{equation}
\begin{split}
    \widehat{\Pr}(T^{\underline{a}_{0}=a}\geq \tau)&=\frac{1}{n}\sum_{i\in \mathcal{C}_0}\widehat{\Pr}(T^{\underline{a}_{0}=a}_{i}\geq \tau|L_{0,i})
    \end{split}
    \label{eq:surv.msm.std}
\end{equation}
where $\Pr(T^{\underline{a}_{0}=a}_{i}\geq \tau|L_{0,i})$ is estimated using the relation $\Pr(T^{\underline{a}_{0}=a}_{i}\geq \tau|L_{0,i})=\exp\left\{-\int_{0}^{\tau}h_{T^{\underline a_0=a}}(u|L_{0,i})du\right\}$, and where the integrals are estimated as outlined above for (\ref{eq:surv.cox.msm}) or (\ref{eq:surv.aalen.msm}). In practice, the empirical  standardization can be done by creating two copies of each individual in $\mathcal{C}_0$ and setting $A_0=A_1=\ldots=A_{\lfloor \tau \rfloor}=1$ for one copy and $A_0=A_1=\ldots=A_{\lfloor \tau \rfloor}=0$ for the other. The covariate values $L_{0,i}$ are the same for both copies. The estimated conditional survival probabilities are then obtained for each individual under both treatment regimes (i.e. for both copies), and then the average calculated across individuals under both treatment regimes.

If it is of interest to estimate the causal risk difference in (\ref{eq:estimand}) for several time horizons $\tau$, it is recommended to fit the MSM using all event times up to $\tau_{\mathrm{max}}$ and with administrative censoring at $\tau_{\mathrm{max}}$ to obtain risk difference estimates for all horizons $\tau\leq \tau_{\mathrm{max}}$, rather than fitting separate MSMs for different time horizons.

Confidence intervals for the estimated causal risk difference can be obtained by bootstrapping. The weights models and the MSM are fitted in each bootstrap sample, so that the bootstrap confidence intervals capture the uncertainty in the estimation of the weights as well as in the MSM.

\subsection{Analysis using the sequential trials approach}
\label{sec:seqtrial}

In the MSM-IPTW approach each individual contributes to the emulated target trial from visit $k=0$, at which time they meet the eligibility criteria. However, individuals may in fact meet the eligibility criteria for the target trial at more than one visit during the study period. In the sequential trials approach an individual can contribute to an emulated trial starting from any visit $k=0,1,2,\ldots$ at which they meet the eligibility criteria. The emulated trial eligibility criteria are applied at each visit $k=0,1,\ldots$ to form a sequence of emulated trials. Recall that individuals who have previously used the treatment are not eligible and so all individuals eligible for the $k$th trial have $\bar A_{k-1}=0$. Those eligible for the $k$th trial therefore include `initiators' who start treatment at visit $k$ ($A_k=1$) and `non-initiators' who remain untreated at visit $k$ ($A_k=0$). Non-initiators in trial $k$ can appear as initiators in a later trial ($k+1,\ldots$). Individuals can therefore appear as initiators in only one trial, but as non-initiators in several trials. 

In this paper we focus on a sequential trials approach in which the length of possible follow-up decreases for trials starting at later visits. An alternative would be to use a sequence of trials that all have equal length of follow-up. To explain this further, consider an example in which the observational cohort has visits at times $k=0,1,2,3,4$ and follow-up to time $5$ but our maximum time horizon of interest for the causal risk difference (\ref{eq:estimand}) is $\tau_{\mathrm{max}}=3$. We could have 5 trials starting at times $k=0,1,2,3,4$, with a decreasing length of follow-up from the start of each trial, such that the trials starting at times $k=0,1,2$ have follow-up of length 3 or longer (and administrative censoring would be applied after 3 time units of follow-up), and the trials starting at times $k=3,4$ only provide 2 and 1 time units of follow-up respectively. In the alternative we would only make use of trials starting at times $k=0,1,2$, with administrative censoring applied after 3 time units of follow-up. 

The sequential trials approach focuses on comparing survival only under the two treatment regimes: `always treated', in which eligible individuals initiate treatment and then always continue treatment, and `never treated', in which eligible individuals do not initiate treatment at the time of meeting eligibility criteria and remain untreated at all later times. Previous descriptions of the sequential trials approach have described the analysis models used but have not been explicit about the causal interpretation of the model parameters. Here we begin by defining MSMs for populations meeting the emulated trial eligibility criteria at visits $k=0,1,2,\ldots,K$ and explain how these relate to the estimand of interest in (\ref{eq:estimand}). We then outline how the MSMs can be estimated using the observational data under certain assumptions, using the methods described by Hernan et al.\cite{Hernan:2008} and Gran et al.\cite{Gran:2010}.

Consider the trial starting at visit $k$. Let $T^{\underline A_k=a}$ denote the counterfactual event time under a treatment regime in which individuals follow their observed treatments up to and including visit $k-1$ and are then assigned to treatment $a$ ($a=0,1$) from visit $k$ onwards (if they survive to visit $k$). Let $h_{k,T^{\underline A_k=a}}(t-k|\bar{A}_{k-1}=0,\bar{L}_{k})$ denote the hazard at time $t-k$ after visit $k$ under this treatment regime, conditional on the baseline characteristics $L_k$ at the start of trial $k$. The time scale is time since the start of the trial. The conditioning on $\bar{A}_{k-1}=0$ indicates the restriction to individuals who have not initiated treatment prior to the start of the $k$th trial, which is part of the eligibility criteria. MSMs for this hazard using a Cox model and Aalen's additive hazard model are 
\begin{equation}
h_{k,T^{\underline A_k=a}}(t-k|\bar A_{k-1}=0,L_{k})=h_{0k}(t-k)\exp\left\{af(t-k;\beta_{Ak})+ \beta_{Lk}L_{k}\right\}
\label{eq:haz.seq}
\end{equation}
and
\begin{equation}
h_{k,T^{\underline A_k=a}}(t-k|\bar A_{k-1}=0,L_{k})=\alpha_{0k}(t-k)+\alpha_{Ak}(t-k)a+\alpha_{Lk}(t-k)L_{k}.
\label{eq:addhaz.seq}
\end{equation}
Recall that $L_k$ includes any time fixed covariates. The conditioning on $L_k$ could be excluded, but we include it because conditioning on baseline variables is a feature of the previously-described sequential trials analysis methods, which we discuss below. In the Cox MSM (\ref{eq:haz.seq}) the hazards in the treated and untreated could be assumed proportional, $f(t-k ;\beta_{Ak})=\beta_{Ak}$, or we could allow the hazard to depend on duration of treatment, e.g. $af(t-k ;\beta_{Ak})=a\beta_{Ak,0}+a\beta_{Ak,1} (t-k)$. The additive hazards MSM (\ref{eq:addhaz.seq}) naturally incorporates a flexible time-dependent effect of treatment on the hazard. Both models could be extended to include interactions between $A_k$ and $L_k$. The above MSMs are conditional on $L_{k}$ rather than $\bar L_k$, which is an assumption. They could be written as conditional on $\bar L_k$, but it is likely to be appropriate that the hazard for each trial $k$ depends on the same amount of history of $L_k$. For example, if we wish to allow the hazard for each trial $k$ to depend on $L_k$ and $L_{k-1}$ then the first trial would need to start at time $k=1$ instead of $k=0$, unless data on the $L$ were available prior to time 0.

Conditional survival probabilities (conditional on $L_k$ and $T\geq k$) for the counterfactual event times are related to the hazard according to the formula
\begin{equation}
\begin{split}
\Pr(T^{\underline{A}_{k}=a}> k+\tau|&T\geq k,\bar A_{k-1}=0, L_{k})\\
&=\exp\left\{-\int_{k}^{k+\tau}h_{k,T^{\underline A_k=a}}(u-k|\bar A_{k-1}=0,L_{k})du\right\}, \quad \tau\leq \tau_{\mathrm{max}}-k.
\end{split}
\label{eq:surv.seq}
\end{equation}
Below we discuss how marginal survival probabilities, and therefore the causal estimand of interest (marginal risk difference), can be obtained. First, however, we outline how the MSMs in (\ref{eq:haz.seq}) or (\ref{eq:addhaz.seq}), and hence the conditional survival probabilities in (\ref{eq:surv.seq}), can be estimated using the observed data. The MSMs in (\ref{eq:haz.seq}) and (\ref{eq:addhaz.seq}) cannot be estimated directly from the observational data because not all individuals meeting the eligibility criteria at visit $k$ are then `always treated' ($\underline A_k=1$) or `never treated' ($\underline A_k=0$). In the sequential trials approach as described by Hernan et al.\cite{Hernan:2008} and Gran et al.\cite{Gran:2010} individuals are artificially censored at the visit at which they switch from their treatment group at the start of a given trial. In the $k$th trial individuals are censored at the earliest visit $m>k$ such that $A_m\neq A_k$. This results in a modified data set in which all individuals in the trial starting at time $k$ have either $\underline A_k=0$ or $\underline A_k=1$ up to the earliest of their event time, their censoring time, or their artificial censoring time due to treatment switch. The treatment switching is expected to depend on time-dependent covariates that are also associated with the outcome, meaning that the artificial censoring is informative. It is addressed using weights, which we refer to as inverse probability of artificial-censoring weights (IPACW). The IPACW at times $0<s<1$ after the start of trial $k$ are equal to 1. The IPACW at times $l\leq s<l+1$ ($l\geq 1$) after the start of trial $k$ is the inverse of the probability that the individual's treatment status at time $l$ remained the same as their treatment status at the start of the trial, $A_k$, conditional on their observed covariates up to time $l$. The MSMs in (\ref{eq:haz.seq}) and (\ref{eq:addhaz.seq}) are then fitted using weighted regression using the time-dependent IPACW, with $a$ in the hazard model replaced by the observed treatment status $A_k$. The conditioning on baseline covariates $L_k$ in each trial controls for confounding of the association between treatment initiation at time $k$ and the hazard. Estimating the MSMs in this way is valid under the same assumptions of positivity, consistency and conditional exchangeability, as are required for the MSM-IPTW analysis. Further details on the IPACW are provided in the Supplementary Materials (Section A1). Hernan et al.\cite{Hernan:2008} used logistic regression to estimate the IPACW, whereas Gran et al.\cite{Gran:2010} used Aalen's additive hazard model. The IPACW could be fitted separately in each trial or combined across trials. After estimating the MSMs using this approach, the expression in (\ref{eq:surv.seq}) can be used to obtain estimates of conditional survival probabilities. The trial starting at $k=0$ provides estimates of $\Pr(T^{\underline{A}_{0}=a}> \tau|L_{0})$ ($\tau\leq \tau_{\mathrm{max}}$), and the trial starting at $k=1$ provides estimates of $\Pr(T^{\underline{A}_{1}=a}>\tau+1|A_{0}=0,L_{1},T\geq1)$ ($\tau\leq \tau_{\mathrm{max}}-1$), ($a=0,1$), and so on for $k=2,\ldots,K$. These are conditional on covariates measured at the start of the trial, $L_k$.

After estimating the conditional survival probabilities in (\ref{eq:surv.seq}) estimates of marginal survival probabilities $\Pr(T^{\underline{A}_{k}=a}> \tau+k|\bar A_{k-1}=0,T\geq k)$ can be obtained via empirical standardization over the distribution of $L_k$ at the start of each trial, as described for the MSM-IPTW approach in Section \ref{sec:msm}. This enables estimation of marginal risk differences:
\begin{equation}
    \Pr(T^{\underline{a}_0=1}>k+\tau|\bar A_{k-1}=0,T\geq k)-\Pr(T^{\underline{a}_0=0}>k+t|\bar A_{k-1}=0,T\geq k), \qquad \tau\leq \tau_{\mathrm{max}}-k.
    \label{eq:estimand.cond}
\end{equation}
The (true) marginal risk differences for the same time horizon $\tau$ after the start of the trial will differ in general for trials starting at different times $k=0,1,2,\ldots$ (for trials for which $\tau\leq \tau_{\mathrm{max}}-k$). Differences in the true value of the estimand for different trials arise when coefficients of the hazard model ((\ref{eq:haz.seq}) or (\ref{eq:addhaz.seq})) depend on $k$, but also because the marginal risk differences for different $k$ refer to populations with different distributions of baseline characteristics, $L_k$. Marginal risk differences that have the same true value across trials starting at different times $k=0,1,\ldots$ can be estimated, using assumptions about the coefficients of the hazard models and standardisation to a common distribution of baseline covariates $L_k$.

Some or all parameters of the MSMs in (\ref{eq:haz.seq}) and (\ref{eq:addhaz.seq}) could be assumed common across trials (i.e. for all $k$). The common parameters can then be estimated by fitting the hazard models using the observed data combined across trials. It may be reasonable in many settings to assume that the impact of the treatment on the hazard at a given time post-initiation does not depend on the visit $k$ at which treatment was initiated, i.e. $\beta_{Ak}=\beta_A$ and $\alpha_{Ak}(t-k)=\alpha_{A}(t-k)$ for all $k$. Similarly the coefficients for $L_k$ could be assumed constant across trials. Often the underlying time scale $t$ will be in a sense arbitrary. For example, in some settings it will be calendar year, and in others the time since joining a cohort. Provided any elements of time (e.g. age, calendar year) that affect the hazard are included in the set of covariates $L_k$ then the baseline hazard may also be assumed common across trials, i.e. $h_{0k}(t-k)=h_{0}(t-k)$ or $\alpha_{0k}(t-k)=\alpha_{0}(t-k)$. 

Suppose that all coefficients of the MSM in (\ref{eq:haz.seq}) and (\ref{eq:addhaz.seq}) are assumed to be the same across trials (i.e. for all $k$). The MSM would be fitted using the data combined across trials using a weighted regression with time-dependent IPACW, and with adjustment for baseline covariates in each trial, $L_k$. When estimating the marginal risk differences by empirical standardization it should be specified clearly which population the marginal probabilities refer to. We suggest that the population of interest for the marginal risk difference is not the population of individuals used in the combined analysis, since this population is not well defined and many individuals appear in it more than once. A better alternative is to estimate the risk difference for the population $\mathcal{C}_0$ of individuals meeting the target trial eligibility criteria at $t=0$ (visit $k=0$). This would make the marginal risk difference comparable with that estimated from the MSM-IPTW analysis. We note that the causal estimand in expression (\ref{eq:estimand}) is implicitly conditional on no prior treatment, which is part of the target trial eligibility criteria. Another possibility would be to estimate the risk difference for the population of individuals meeting the target trial eligibility criteria at the last visit time $K$, as this population might be expected to be the most similar to the patient population now in terms of their distribution of characteristics. 

Suppose, instead, that we wish to allow the baseline hazard the differ across trials, but that the coefficients for treatment ($a$) and baseline covariates ($L_k$) are assumed the same across trials. The MSMs in(\ref{eq:haz.seq}) or (\ref{eq:addhaz.seq}) could be fitted using data combined across trials, but with a stratified baseline hazard. In this case, care should be taken over which baseline hazard is used to obtain the marginal risk difference estimates. For example, the baseline hazard corresponding to the trial starting at $k=0$ may be used to obtain the marginal risk difference estimates for all time horizons $\tau\leq \tau_{\mathrm{max}}$. This would make the marginal risk difference comparable with that estimated from the MSM-IPTW analysis. The baseline hazard from the trial starting at the penultimate visit $K-1$ can only be used to estimate marginal risk difference estimates for time horizons $\tau\leq \tau_{\mathrm{max}}-(K-1)$, and using baseline hazards corresponding to trials starting at visit $k\geq 1$ to estimate marginal risk differences would result in marginal risk differences that do not correspond to those estimated from the MSM-IPTW analysis. 

In their descriptions of the sequential trials approach Gran et al.\cite{Gran:2010} used the Cox proportional hazards model, and Hernan et al.\cite{Hernan:2008} used pooled logistic regression, which is equivalent as the times between visits gets smaller \citep{D'Agostino:1990}. R{\o}ysland et al.\cite{Roysland:2011} used an additive hazards model, though their aim was to estimate direct and indirect effects, which differs from our aims in this paper. Gran et al.\cite{Gran:2010} assumed the hazard ratios for treatment in the Cox model to be common across the trials, but allowed a different baseline hazard in each trial. Hernan et al.\cite{Hernan:2008} allowed the odds ratio for treatment to differ across trials and also performed a test for heterogeneity, though it is unclear whether they allowed separate intercept parameters (baseline hazards) across trials. If the treatment effect is assumed the same across trials, it is recommended to assess this assumption using a formal test. 

As in the MSM-IPTW approach, confidence intervals for the estimated causal risk difference can be obtained by bootstrapping. The formation of the sequential trials, estimation of the weights, fitting of the conditional MSMs, and obtaining the risk difference are all repeated in each bootstrap sample.


\section{Comparing MSM-IPTW and the sequential trials approach}
\label{sec:comparison}

\subsection{Estimating causal effects: an illustrative example using a causal tree diagram}
\label{sec:est.comp}

In this section we show how the MSM-IPTW and sequential trials approaches can estimate the same causal estimand using an illustrative example and non-parametric estimates. Our aim is to provide insight into how the two approaches use the data differently to estimate the same quantity. 

Consider a situation as depicted in the DAG in Figure \ref{fig:dag}, but with treatment and covariate information only collected at up to two visits ($L_0,A_0$ measured at time 0, and $L_1,A_1$ measured at time 1) and survival status observed at times 1 ($Y_1=I(0\leq T<1)$) and 2 ($Y_2=I(1\leq T< 2)$). We focus on a single binary time-dependent confounder $L$ and omit the unobserved variable $U$. Figure \ref{fig:tree} shows a causal tree graph (see for example Richardson and Rotnitzky\cite{Richardson:2014}), representing the different possible combinations of the variables $L_0,A_0,Y_1$ up to time 1, and then the different combinations of $L_1,A_1,Y_2$ among individuals who survive to time 1 ($Y_1=0$). A total of $n$ individuals are observed at time 0. The numbers in brackets on the branches of the tree are the number of individuals who followed a given pathway up to that branch. The notation is as follows: $n_{l_0}$ denotes the number with $L_0=l_0$; $n_{l_0,a_0}$ the number with $L_0=l_0,A_0=a_0$; $n^{y_1}_{l_0a_0}$
the number with $L_0=l_0,A_0=a_0, Y_1=y_1$; $n^{0}_{l_0a_0,l_1}$
the number with $L_0=l_0,A_0=a_0, Y_1=0, L_1=l_1$; $n^{0}_{l_0a_0,l_1a_1}$
the number with $L_0=l_0,A_0=a_0, Y_1=0, L_1=l_1,A_1=a_1$; and $n^{0y_2}_{l_0a_0,l_1a_1}$
the number with $L_0=l_0,A_0=a_0, Y_1=0, L_1=l_1,Y_2=y_2$.

Our interest is in comparing survival probabilities under two treatment regimes: always treated ($A_0=A_1=1$) and never treated ($A_0=A_1=0$). The branches representing these two treatment regimes are shown with thick lines in the causal tree diagram. The MSM analysis makes use of the causal tree diagram as depicted in Figure \ref{fig:tree}. Under the sequential trials approach, we create a trial starting at time 0 and a trial starting at time 1. In the trial starting at time 0 individuals who survive to time 1 are censored at time 1 if $A_1\neq A_0$, i.e. not all branches after time 1 are used. The trial starting at time 1 is restricted to individuals with $A_0=0$ and $Y_1=0$. The sections of the causal tree diagram for these two trials are shown in Figure \ref{fig:tree.seq}. 

Consider the probability of survival to time 1 with treatment $a$, $\Pr(Y_1^{A_0=a}=0)$. The MSM-IPTW approach estimates this using the results that (under the conditions of consistency, positivity and conditional exchangeability)
\begin{equation}
    \Pr(Y_1^{A_0=a}=0)=E\left\{\frac{I(A_0=a)(1-Y_1)}{\Pr(A_0=a|L_0)}\right\}.
\end{equation}
Based on the tree diagram, a non-parametric estimate of this is
\begin{equation}
    \widehat{\Pr}(Y_1^{A_0=a}=0)=\frac{1}{n}\left\{n_{0a}^{0}\left(\frac{n_{0a}}{n_0}\right)^{-1}+n_{1a}^{0}\left(\frac{n_{1a}}{n_1}\right)^{-1}\right\},
    \label{eq:msm.tree.1}
\end{equation}
where the two contributions come from individuals with $L_0=0$ and $L_0=1$. On the other hand, the sequential trial starting at time 0 estimates $\Pr(Y_1^{A_0=a}=0|L_0)$. By consistency and conditional exchangeability we have $\Pr(Y_1^{A_0=a}=0|L_0)=\Pr(Y_1=0|A_0=a,L_0)$.
Based on the tree diagram, a non-parametric estimate of this using the trial starting at time 0 is $\Pr(Y_1=0|A_0=a,L_0=l_0)=\frac{n_{l_0 a}^0}{n_{l_0 a}}, \quad l_0=0,1$. Using the result that $\Pr(Y_1^{A_0=a}=0)=\sum_{l_0}\Pr(Y_1=0|A_0=a,L_0=l_0)\Pr(L_0=l_0)$ (assuming consistency and conditional exchangeability) it follows that a non-parametric estimate of the marginal probability $\Pr(Y_1^{A_0=a}=0)$ based on the trial starting at time 0 is
\begin{equation}
    \widehat{\Pr}(Y_1^{A_0=a}=0)=\frac{n_{0 a}^0}{n_{0 a}}\frac{n_{0}}{n}+\frac{n_{1 a}^0}{n_{1 a}}\frac{n_{1}}{n},
\end{equation}
which is the same as the estimate using MSM-IPTW (i.e. equation (\ref{eq:msm.tree.1})). Therefore the MSM-IPTW and sequential trials approaches give identical estimates of $\Pr(Y_1^{A_0=a}=0)$. The equivalence between inverse probability weighted estimates and standardized estimates obtained using the g-formula in the non-parametric setting is well established (see for example \cite{Daniel:2013,HernanRobins:2010}). In the sequential trials approach, the trial starting at time 1 can be used to estimate the probability of survival to time 2 after initiating the treatment strategy ($a=0,1$) conditional on survival to time 1 and on $A_1=0$. The non-parametric estimate is
\begin{equation}
\begin{split}
    \widehat{\Pr}(Y_2^{ A_1=a}=0|Y_1=0,A_1=0)=&\left(\frac{n_{00,0a}^{0,0}+n_{10,0a}^{0,0}}{n_{00,0a}^{0}+n_{10,0a}^{0}}\right)\left(\frac{n_{00,0}^{0}+n_{10,0}^{0}}{n_{00}^{0}+n_{10}^{0}}\right)\\
    &+\left(\frac{n_{00,1a}^{0,0}+n_{10,1a}^{0,0}}{n_{00,1a}^{0}+n_{10,1a}^{0}}\right)\left(\frac{n_{00,1}^{0}+n_{10,1}^{0}}{n_{00}^{0}+n_{10}^{0}}\right).
    \end{split}
\end{equation}
The marginal probability estimate $\widehat{\Pr}(Y_1^{A_0=a}=0)$ refers to a population in which $\Pr(L_0=1)=n_1/n$, whereas the marginal probability estimate $\widehat{\Pr}(Y_2^{ A_1=a}=0|Y_1=0,A_1=0)$ refers to a population with a different distribution of $L_1$. The estimate from the trial starting at time 1 could alternatively be standardized to the distribution of $L_0$ at time 0:
\begin{equation}
    \widehat{\Pr}^{\mathrm{Std}}(Y_2^{ A_1=a}=0|Y_1=0,A_1=0)=\left(\frac{n_{00,0a}^{0,0}+n_{10,0a}^{0,0}}{n_{00,0a}^{0}+n_{10,0a}^{0}}\right)\left(\frac{n_{0}}{n}\right)+\left(\frac{n_{00,1a}^{0,0}+n_{10,1a}^{0,0}}{n_{00,1a}^{0}+n_{10,1a}^{0}}\right)\left(\frac{n_{1}}{n}\right).
\end{equation}
Under the assumption that $\Pr^{\mathrm{Std}}(Y_2^{ A_1=a}=0|Y_1=0,A_1=0)=\Pr(Y_1^{A_0=a}=0)$, a combined estimate of $\Pr(Y_1^{A_0=a}=0)$ could be obtained from the two trials, for example using an inverse-variance-weighted combination of the estimates from the two trials.

We can also show that non-parametric estimates of $\Pr(Y_2^{\bar A_1=a}=0)$ from the two methods are the same. The MSM-IPTW approach uses 
\begin{equation}
    \Pr(Y_2^{\bar A_1=a}=0)=E\left\{\frac{I(A_0=a,A_1=a)(1-Y_1)(1-Y_2)}{\Pr(A_0=a|L_0)\Pr(A_1=a|A_0=a,L_0,L_1)}\right\}.
\end{equation}
Based on the tree diagram, a non-parametric estimate of this is
\begin{equation}
\begin{split}
    \widehat{\Pr}(Y_2^{\bar A_1=a}=0)=&\frac{1}{n}\left\{n_{0a,0a}^{0,0}\left(\frac{n_{0a}}{n_0}\right)^{-1}\left(\frac{n_{0a,0a}^{0}}{n_{0a,0}^{0}}\right)^{-1}+
    n_{0a,1a}^{0,0}\left(\frac{n_{0a}}{n_0}\right)^{-1}\left(\frac{n_{0a,1a}^{0}}{n_{0a,1}^{0}}\right)^{-1}\right.\\
    &+\left.n_{1a,0a}^{0,0}\left(\frac{n_{1a}}{n_1}\right)^{-1}\left(\frac{n_{1a,0a}^{0}}{n_{1a,0}^{0}}\right)^{-1}+
    n_{1a,1a}^{0,0}\left(\frac{n_{1a}}{n_1}\right)^{-1}\left(\frac{n_{1a,1a}^{0}}{n_{1a,1}^{0}}\right)^{-1}\right\},
    \end{split}
    \label{eq:msm.nonpara.2}
\end{equation}
where the four contributions come from individuals with the four possible combinations of $(L_0,L_1)$. 

The sequential trials approach estimates $\Pr(Y_2^{\bar A_1=a}=0|L_0)$, which can be written
\begin{equation}
    \Pr(Y_2^{\bar A_1=a}=0|L_0)=\Pr(Y_2^{\bar A_1=a}=0|Y_1^{A_0=a}=0,L_0)\Pr(Y_1^{A_0=a}=0|L_0).
\end{equation}
The term $\Pr(Y_1^{A_0=a}=0|L_0)$ was considered above. The first term, $\Pr(Y_2^{\bar A_1=a}=0|Y_1^{A_0=a}=0,L_0)$, is estimated by IPACW, because individuals with $A_1\neq A_0$ are censored at time 1, and the remaining individuals with $A_1=A_0$ weighted by $\Pr(A_1=a|A_0=a,L_0,L_1)^{-1}$. Under the assumptions of consistency, positivity and conditional exchangeability we can write 
\begin{equation}
    \Pr(Y_2^{\bar A_1=a}=0|Y_1^{A_0=a}=0,L_0)=E\left\{\frac{I(A_1=a)(1-Y_2)}{\Pr(A_1=a|A_0=a,L_0,L_1)}|Y_1^{A_0=a}=0,A_0=a,L_0 \right\}.
\end{equation}
Based on the tree diagram, a non-parametric estimate of this is
\begin{equation}
\begin{split}
    \widehat{\Pr}(Y_2^{\bar A_1=a}=0|Y_1^{A_0=a}=0,L_0=l_0)&=\widehat{\Pr}(Y_2^{\bar A_1=a}=0|Y_1=0,A_0=a,L_0=l_0)\\
    &=\frac{1}{n_{l_0 a}^{0}}\left\{n_{l_0 a,0a}^{0,0}\left(\frac{n_{l_0 a,0a}^{0}}{n_{l_0 a,0}^{0}}\right)^{-1}+n_{l_0 a,1a}^{0,0}\left(\frac{n_{l_0 a,1a}^{0}}{n_{l_0 a,1}^{0}}\right)^{-1} \right\}.
    \end{split}
    \label{eq:seq.nonpara.2}
\end{equation}
Using this result along with the estimate $\widehat{\Pr}(Y_1^{A_0=a}=0|L_0=l_0)=\widehat{\Pr}(Y_1=0|A_0=a,L_0=l_0)=n_{l_0a}/n_{l_0}$, it can be shown that the sequential trials estimate of $\Pr(Y_2^{\bar A_1=a}=0)=\sum_{l_0=0,1}\Pr(Y_2^{\bar A_1=a}=0|L_0=l_0)\Pr(L_0=l_0)$ is the same as the MSM-IPTW estimate in (\ref{eq:msm.nonpara.2}). By comparing (\ref{eq:msm.nonpara.2}) and (\ref{eq:seq.nonpara.2}) we can see clearly the connection between the weights used in the MSM-IPTW approach and those used in the sequential trials approach (IPACW). 

\subsection{Specification and estimation of MSMs}

In practice, model-based estimates are typically required instead of non-parametric estimates, as we usually have multiple time-dependent confounders to consider, including continuous variables. In this section we compare in more detail the form of the MSMs used in the two approaches, and how they are estimated using inverse probability weights.  


A key difference in the MSMs used in the two approaches is that in the MSM-IPTW the MSM for the hazard at time $t$ includes the history of treatment up to time $t$, $\bar A_{\lfloor t \rfloor}$, whereas the MSM used in the sequential trials approach involves only the treatment status assigned at the start of the trial. 
The MSM-IPTW approach therefore requires that the relation between treatment history $\bar A_{\lfloor t \rfloor}$ and the hazard is correctly specified, whereas in the sequential trials approach there are limited options for the form of the MSM because only two treatment patterns are considered, because after the artificial censoring all individuals used in the analysis are either always treated or never treated. In MSM-IPTW, when baseline covariates are excluded from the MSM, the MSM is a fully marginal model - the possibility of interaction between treatment and baseline covariates is not ruled out but does not have to be modelled. If the MSM used in the MSM-IPTW approach includes baseline covariates $L_0$ then any interactions that exist between treatment and $L_0$ must be included in the MSM and correctly specified. Similarly, in the sequential trials approach, if there are interactions between treatment and the baseline covariates at the start of each trial, $L_k$, then these must be included in the MSM and correctly specified. MSMs that condition on baseline covariates are therefore more susceptible to misspecification. The sequential trials approach could use weighting in the first time period instead of adjustment for baseline covariates, though that is not how the method has been described or used to date. Conditioning on baseline covariates in the sequential trials approach enables use of empirical standardization to obtain risk difference estimates for a population with any distribution of the baseline covariates. When combined with assumptions about the form of the MSM for the hazard in each trial, this enables us to estimate the same causal estimand in the sequential trials approach as in the MSM-IPTW approach.

If the sequential trials analysis makes the assumption that the effect of treatment on the hazard is the same in all trials, i.e. does not depend on the time of treatment initiation, then the sequential trials approach has information about the effect of treatment initiation from several time origins. 
If the assumptions are valid, this re-use of individuals from several time origins could lead to gains in efficiency in the marginal risk difference estimate at some time horizons $\tau$ relative to the MSM-IPTW approach. On the other hand, in the MSM-IPTW analysis individuals who switch their treatment status during follow-up continue to contribute information to the analysis, whereas in the sequential trials analysis individuals cease to contribute information when they deviate from their treatment group at the start of a given trial. This means that in the sequential trials analysis the number of individuals with longer term follow-up will be smaller than in the MSM-IPTW analysis. 

To fit the MSMs, both approaches require time-updated weights, and hence the data should be formatted with multiple rows per individual (one for each visit for MSM-IPTW, and one for each visit within each trial for the sequential trials approach). In the MSM-IPTW approach (without conditioning on baseline covariates), the MSM is estimated by weighting individuals in the observed data using time-updated weights. In the sequential trials approach the MSM is instead estimated through adjustment in trial $k$ for baseline covariates $L_{k}$ and artificial censoring of individuals when they deviate from their initial treatment group at time $k$. The artificial censoring is addressed using IPACW. The weight up to time 1 after the start of a given trial is always equal to 1. In the MSM-IPTW approach with conditioning on baseline variables $L_0$ the IPTW weight at a given time is the same as the IPACW weight used in the sequential trials approach for the trial starting at time 0 for individuals following the `always treated' or `never treated' regimes. In the MSM-IPTW approach without conditioning on baseline variables the IPTW weights are proportional to the IPACW weights used in the sequential trials approach. Because the MSM-IPTW approach includes individuals following any treatment regime (not just `always treated' or `never treated'), we may expect to see more extreme weights with increasing follow-up, compared with the weights used in the sequential trials approach.


Because the two approaches use the data differently and because the MSM-IPTW approach could make use of more extreme weights, it is of interest to investigate the relative efficiency of the two approaches for estimating the causal estimand. We note that it is only appropriate to consider the relative efficiency of the two approaches when they target the same causal estimand, which, as discussed in section \ref{sec:seqtrial}, requires assumptions that some parameters of the MSM used in the sequential trials approach are common across trials.


\subsection{Compatibility of MSMs used in the two approaches}
\label{sec:model.based}

\begin{figure}
    \centering
        \caption{Causal tree diagram illustrating a study with a binary treatment $A_k$ and binary covariate $L_k$, both measured at two time points ($t=0,1$). $Y_t$ is a discrete time survival outcome. Thick lines indicate branches for groups of individuals who were untreated or treated at both time points.}
    \includegraphics[scale=0.6]{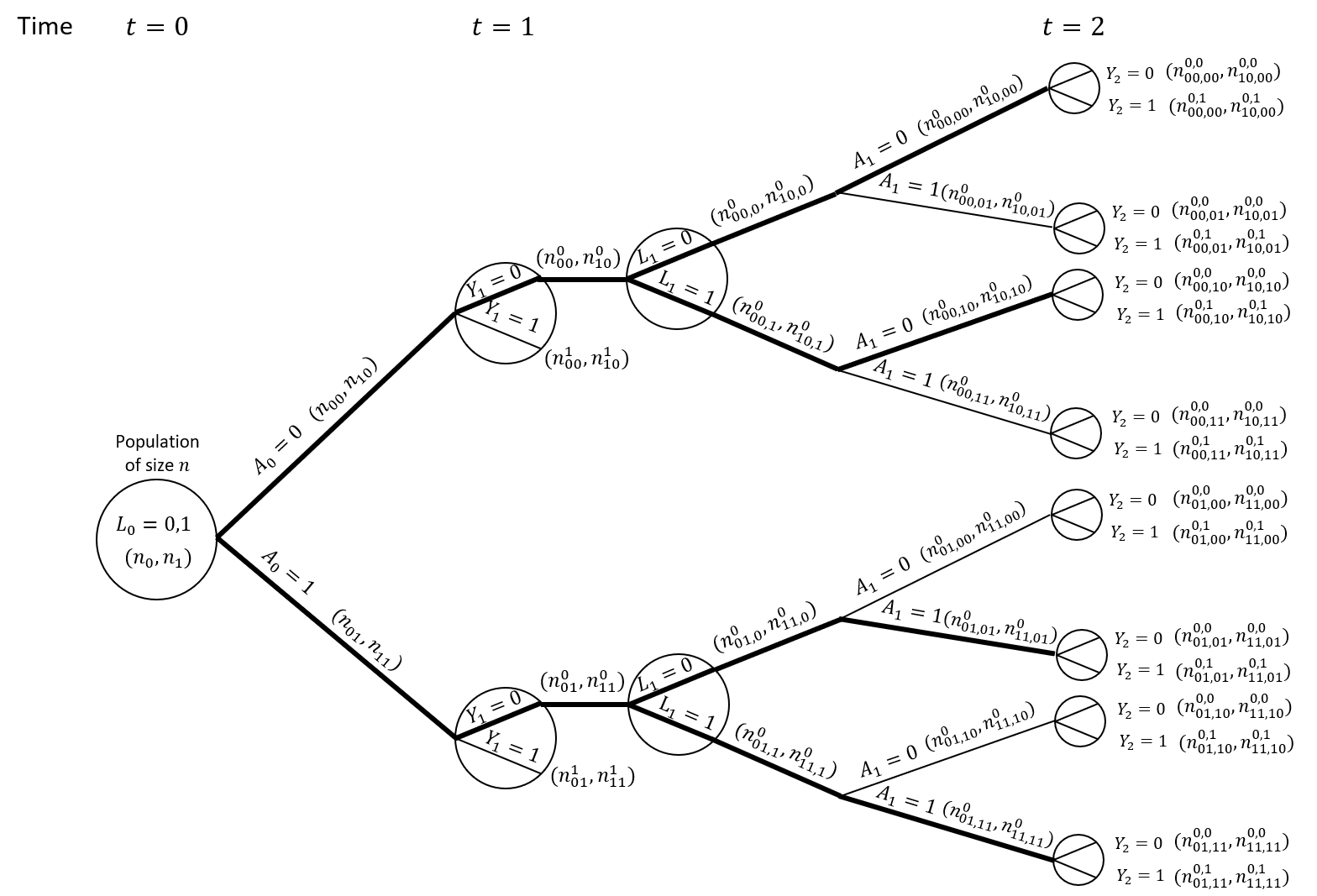}
    \label{fig:tree}
\end{figure}

\begin{figure}
    \centering
        \caption{Illustration of the sequential trials approach using causal tree diagram from Figure \ref{fig:tree}. (a) In trial starting at time $t=0$ individuals with $Y_1=0$ are censored at time 1 if $A_1\neq A_0$. (b) The trial starting at time $t=1$ is restricted to individuals with $A_0=0$ (and $Y_1=0$).}
    \includegraphics[scale=0.85]{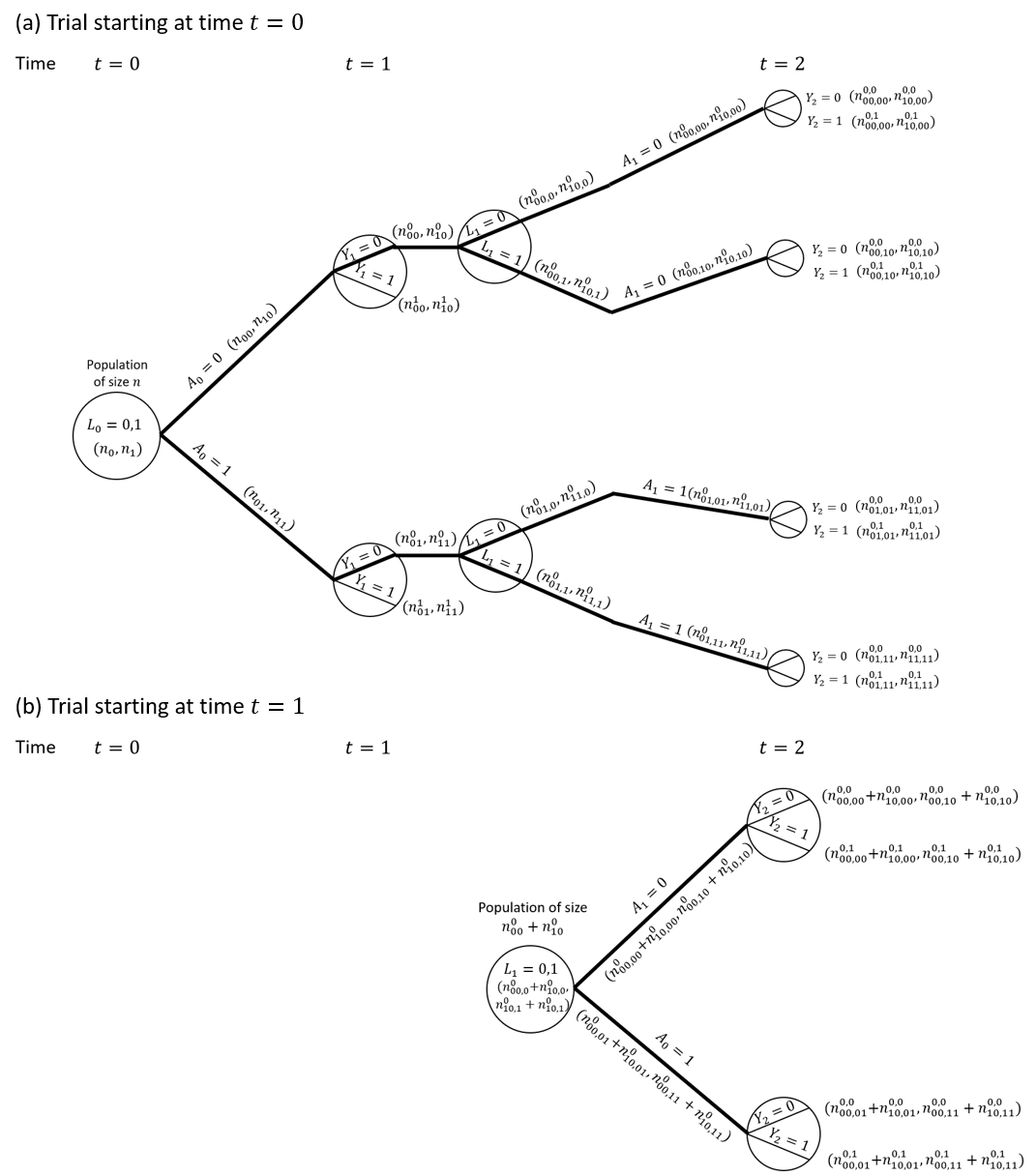}
    \label{fig:tree.seq}
\end{figure}

In Sections \ref{sec:msm} and \ref{sec:seqtrial} we considered MSMs based on Cox models or on additive hazard models. In this section we show that one can use an additive hazard model for the MSM for the sequential trials analysis (expression (\ref{eq:addhaz.seq})) (with common parameters assumed across trials) and an additive hazard model for the MSM used in the MSM-IPTW approach (expression (\ref{eq:msm.aalen}), without conditioning on $L_0$), and that both MSMs can be correctly specified. However, if we use a Cox model for the MSM for the sequential trials analysis (expression (\ref{eq:haz.seq})) and a Cox model for the MSM used in the MSM-IPTW approach (expression (\ref{eq:msm.cox}), then both MSMs cannot simultaneously be correctly specified in general. This is because the parameters of additive hazard models are collapsible, whereas hazard ratios in the Cox model are non-collapsible. 
Martinussen and Vansteelandt (2013)\cite{Martinussen:2013} explained the implications of collapsibility for the use of Aalen additive hazards models and Cox models in the context of estimating the causal effect of a point treatment on survival. Keogh et al. (2021)\cite{Keogh:2021} extended to a longitudinal setting similar to that considered in this paper. Here we outline some key points in the context of the simple setting depicted in Figures \ref{fig:tree} and \ref{fig:tree.seq}. 

For the MSM-IPTW analysis in this section we again focus on an MSM that is not conditional on $L_0$, as in (\ref{eq:msm.cox}),(\ref{eq:msm.aalen}). The MSMs used in a MSM-IPTW analysis and in the sequential trials analysis may be considered to be consistent with each other if there exists an underlying fully conditional hazard model that implies both the MSM used in MSM-IPTW and the MSM used in the sequential trials approach, which is conditional on baseline covariates at the start of each trial. As above we consider estimating $\Pr(Y_1^{A_0=a}=0)$ (or $\Pr(T^{\underline A_0=a}>1)$) and $\Pr(Y_2^{\bar A_1=a}=0)$ (or $\Pr(T^{\underline A_1=a}>2)$). First consider a fully conditional additive hazard model of the form 
\begin{equation}
h(t|\bar A_{\lfloor t \rfloor},\bar L_{\lfloor t \rfloor}) =  \alpha_0(t)+\alpha_A(t) A_{\lfloor t \rfloor}+\alpha_L(t) L_{\lfloor t \rfloor}.
\label{eq:haz.fullcond}
\end{equation}
Under this model the conditional survival probability at time 1 is 
\begin{equation}
    \Pr(Y_1=0|A_0,L_0)=\exp\left(-\int_{0}^{1}(\alpha_0(u)+\alpha_A(u) A_0+\alpha_L(u) L_0) du\right)
\end{equation}
 and the marginal probability of survival to time 1 is
\begin{equation}
\begin{split}
\Pr(Y_1^{A_0=a}=0)&=e^{-\int_{0}^{1}(\alpha_0(u)+\alpha_A(u) A_0) du}\Pr(L_0=0)+e^{-\int_{0}^{1}(\alpha_0(u)+\alpha_A(u) A_0+\alpha_L(u)) du}\Pr(L_0=1)\\
&=e^{-\int_{0}^{1}(\alpha_0^*(u)+\alpha_A(u) A_0) du},
\end{split}
\end{equation}
where $\int_{0}^{1}\alpha_0^*(u)=\int_{0}^{1}\alpha_0(u)du+\log\{\Pr(L_0=0)+e^{-\int_{0}^{1}\alpha_L(u)du}\Pr(L_0=1)\}$. This expression is in the form of the survival probability from a marginal additive hazard model of the form 
\begin{equation}
h_{T^{^{A_0=a}}}(t) =  \alpha_0^*(t)+\alpha_A(t) a.
\end{equation}
It follows that we could use an additive hazard model for $\Pr(Y_1^{A_0=a}=0|L_0)$ in the sequential trials analysis and an additive hazard model for $\Pr(Y_1^{A_0=a}=0)$ in the MSM-IPTW analysis, and that both models can be correctly specified.

Next consider the probabilities $\Pr(Y_2^{\bar A_1=a}=0)=\Pr(Y_2^{\bar A_1=a}=0|Y_1^{A_0=a}=0)\Pr(Y_1^{A_0=a}=0)$ and $\Pr(Y_2^{\bar A_1=a}=0|L_0)=\Pr(Y_2^{\bar A_1=a}=0|Y_1^{A_0=a}=0,L_0)\Pr(Y_1^{A_0=a}=0|L_0)$. Under the fully conditional additive hazard model in (\ref{eq:haz.fullcond}) it can be shown that (see Supplementary Materials Section A3)
\begin{equation}
     \Pr(Y_2^{\bar A_1=a}=0|Y_1^{A_0=a}=0,L_0)=\exp\left\{-\int_1^{2}(\alpha_0(u)+\alpha_(u)a)du+\Delta(a,L_0)\right\}
\end{equation}
and 
\begin{equation}
     \Pr(Y_2^{\bar A_1=a}=0|Y_1^{A_0=a}=0)=\exp\left\{-\int_1^{2}(\alpha_0(u)+\alpha_(u)a)du+\Delta^*(a)\right\},
\end{equation}
where 
$$
\Delta(a,L_0)=\log\left\{\Pr(L_1=0|Y_1=0,A_0=a,L_0)+e^{-\int_1^2\alpha_L(u)du}\Pr(L_1=1|Y_1=0,A_0=a,L_0)\right\}
$$
and
$$
\Delta^*(a)=\log\left\{\frac{e^{\Delta(a,0)}\Pr(L_0=0)+e^{\Delta(a,0)-\int_0^1\alpha_L(u)du}\Pr(L_1=1)}{\Pr(L_0=0)+e^{-\int_0^1\alpha_L(u)du}\Pr(L_1=1)}\right\}.
$$
It follows that there exists an underlying conditional hazard model that gives rise to an additive model for the sequential trials analysis and for the MSM-IPTW analysis. 

Secondly consider a fully conditional Cox proportional hazards model of the form 
\begin{equation}
h(t|\bar A_{\lfloor t \rfloor},\bar L_{\lfloor t \rfloor}) =  h_0(t)e^{\beta_A A_{\lfloor t \rfloor}+\beta_L L_{\lfloor t \rfloor}}.
\end{equation}
Under this model the conditional survival probability at time 1 is $\Pr(Y_1=0|A_0,L_0)=e^{-H_0(1)e^{\beta_A A_0+\beta_L L_0}}$ where $H_0(1)=\int_{0}^{1}h_0(u)du$. Using this, the marginal probability of interest can be written 
\begin{equation}
\Pr(Y_1^{A_0=a}=0)=\exp\left(-H_0(1)e^{\beta_A A_0}\right)\Pr(L_0=0)+\exp\left(-H_0(1)e^{\beta_A A_0+\beta_L }\right)\Pr(L_0=1).
\end{equation}
This expression for $\Pr(Y_1^{A_0=a}=0)$ is not the form of the survival probability from a marginal Cox proportional hazards model. Therefore, if a Cox model was assumed for $h_{T^{\underline{a}_{0}}}(t|L_0)$ in the sequential trials analysis, then this does not imply a Cox model for $h_{T^{\underline{a}_{0}}}(t)$ in the MSM-IPTW analysis, and vice-versa. If Cox models are used for a sequential trials analysis and a MSM-IPTW analysis then the two analyses make different modelling assumptions that cannot both be true. In practice, however, a Cox model could be a good working model for both approaches. 

The result in this section has implications for comparing estimates obtained from the MSM-IPTW approach and the sequential trials approach, and we make use of this in the simulation study below by using additive hazards models. 

\section{A simulation study for the comparison of MSM-IPTW and the sequential trials approach}
\label{sec:sim}

\subsection{Simulation plan}

To evaluate and compare the performance of the two methods in question we will now consider a simulation study based on the data generating simulation algorithm for longitudinal and time-to-event data described by Keogh et al. \cite{Keogh:2021}. We follow the general recommendations of Morris et al.(2019)\cite{Morris:2019} on the conduct of simulation studies for evaluating statistical methods. R code for replicating the simulation is provided at https://github.com/ruthkeogh/sequential\_trials.

\emph{Aim}

Our aim is to compare the MSM-IPTW and sequential trials approaches for estimating the effect of a time-varying treatment on survival, subject to time-dependent confounding, from longitudinal observational data. Our focus is on a setting in which both approaches target the same causal estimand (see below) and hence it is relevant to assess their relative efficiency. Both methods are expected to produce approximately unbiased estimates when the modelling assumptions are met. We hypothesise that the sequential trials analysis could be more efficient than the MSM-IPTW analysis in certain settings, depending on the data generating mechanism and time horizon $\tau$. MSM-IPTW is inefficient when there are extreme weights. The efficiency of the sequential trials approach is expected to depend on the proportion of individuals always in the treatment group or always in the control group, since when many individuals switch over time this would result in few people contributing as follow-up time increases, and the potential for large weights.

\emph{Data generating mechanisms}

Data are generated according to the DAG in Figure 1. Individuals are observed at up to 5 visits, with administrative censoring at time 5. We consider three simulation scenarios, starting with a `standard' scenario (scenario 1) and then investigating the performance of the methods when certain aspects of the data generating mechanism are varied. The simulation scenarios are outlined in detail in Table \ref{tab:sim.scen}. In all scenarios, once an individual starts treatment they remain on treatment, i.e., for any $k$, there are no individuals that go from $A_k=1$ to $A_{k+1}=0$. Also in all scenarios there is a single time-dependent confounder $L$, the model for $L_k$ conditional on $A_{k-1}$, $L_{k-1}$,$k$ and $U$ is the same, and event times are generated according to the same conditional additive hazards model
\begin{equation}
h(t|\bar A_{\lfloor t \rfloor},\bar{L}_{\lfloor t \rfloor},U)=\alpha_{0}+\alpha_{A}A_{\lfloor t \rfloor}+\alpha_{L}L_{\lfloor t \rfloor}+\alpha_U U.
\label{eq:sim.condhaz}
\end{equation}
In scenarios 1 and 2 the log odds ratio for $L_k$ in the logistic model for the probability of $A_k=1$ is 0.5, giving `moderate' dependence of treatment initiation on $L_k$. In scenario 3, the log odds ratio for $L_k$ in the logistic model for the probability of $A_k=1$ is increased to 3, meaning that there is strong dependence of treatment initiation on $L_k$.

In scenario 1 the intercept in the logistic model for the probability of $A_k=1$ is such that 25\% of individuals have $A_0=1$. In scenario 2 the intercept in the logistic model for the probability of $A_k=1$ is reduced so that the proportion of individuals initiating treatment at a given visit is lower, with approximately 5\% having $A_0=1$.

In scenarios 1, 2 and 3, respectively, approximately 57\%, 62\% and 56\% of individuals have the event of interest during the 5 years of follow-up. We do not include any censoring apart from administrative censoring, though extensions to include this are straightforward.

The reason for using an additive hazards model for generating event times is that it enables us to specify a model that is correct under both the MSM-IPTW and the sequential trials analysis, as discussed in Section \ref{sec:model.based}, thus enabling a fair comparison between the two approaches. 

\emph{Estimand}

The estimand of interest is the marginal risk difference in (\ref{eq:estimand}). We consider time horizons on a continuous scale up to time $\tau=5$. The population of interest for the marginal risk difference is the set of $n$ individuals observed at time 0.

\emph{Methods}

The data are analysed using the MSM-IPTW approach and the sequential trials approach. In the MSM-IPTW approach we considered MSMs with and without conditioning on $L_0$. When the conditional additive hazard model is as in (\ref{eq:sim.condhaz}), it can be shown \citep{Keogh:2021} that the correct form for the MSM used in the MSM-IPTW analysis using the MSM that is not conditional on $L_0$ is
\begin{equation}
h_{T^{\underline{a}_{0}}}(t)=\alpha_{0}(t)+\sum_{j=0}^{\lfloor t \rfloor}\tilde{\alpha}_{Aj}(t)a_{\lfloor t\rfloor -j}
\label{eq:msm.sim1}
\end{equation}
and the correct form for the MSM conditional on $L_0$ is
\begin{equation}
h_{T^{\underline{a}_{0}}}(t|L_0)=\alpha_{0}(t)+\sum_{j=0}^{\lfloor t \rfloor}\tilde{\alpha}_{Aj}(t)a_{\lfloor t\rfloor -j}+\alpha_L(t) L_0.
\label{eq:msm.sim2}
\end{equation}
The correct form for the MSM used in the sequential trials analysis, under our data generating mechanism, is
\begin{equation}
h_{T^{\underline{a}_{k}=a}}(t-k|L_k)=\alpha_{0}(t-k)+\alpha_{A}(t-k)a+\alpha_{L}(t-k)L_k, \qquad k=0,\ldots,4.
\end{equation}
The coefficients in the fully conditional hazard model in (\ref{eq:sim.condhaz}) do not depend on time, and this results in the coefficients in the MSM for the sequential trials analysis being a function of time since the start of the trial, but being the same across trials $k=0,\ldots,4$.

We use stabilized weights for all analyses. In the MSM-IPTW approach using the MSM in (\ref{eq:msm.sim1}) (without conditioning on $L_0$) the stabilized IPTW at time $t$ are
\begin{equation}
    \prod_{k=0}^{\lfloor t\rfloor}\frac{\Pr(A_k|\bar A_{k-1})}{\Pr(A_k|\bar L_k,\bar A_{k-1})}
\end{equation}
and in the MSM-IPTW approach using the MSM in (\ref{eq:msm.sim2}) (with conditioning on $L_0$) the stabilized IPTW at time $t$ are
\begin{equation}
    \prod_{k=0}^{\lfloor t\rfloor}\frac{\Pr(A_k|\bar A_{k-1},L_0)}{\Pr(A_k|\bar L_k,\bar A_{k-1})},
\end{equation}
 In the sequential trials approach the stabilized IPACW at time $t$ (for $t\geq 1$) after the start of the trial for the trial starting at visit $k$ are (for $t\geq 1$)
\begin{equation}
\left\{\begin{array}{cc}
   1  & \mbox{for }A_k=1 \\
   \prod_{j=1}^{\lfloor t\rfloor}\frac{\Pr(A_{k+j}=0|L_k,\bar{A}_{k+j-1}=0)}{\Pr(A_{k+j}=0|\bar L_j,\bar{A}_{k+j-1}=0)}  & \mbox{for }A_k=0
\end{array}\right.
\label{eq:wt.seq}
\end{equation}
The IPACW are equal to 1 up to time 1 after the start of each trial, i.e. up to time $k+1$ for the trial starting at visit $k$.

The weights take into account that individuals do not stop treatment once they start. In MSM-IPTW we have $\Pr(A_k=1|\bar A_{k-1}=1)=1$, $\Pr(A_k=1|\bar L_k,\bar A_{k-2},A_{k-1}=1)=1$ and $\Pr(A_k=1|\bar A_{k-1}=1,L_0)=1$. All other probabilities used in the weights were estimated using logistic regression models fitted using all visits combined ($k=0,\ldots,4$). The probabilities $\Pr(A_k|\bar A_{k-2},A_{k-1}=0)$ were estimated using a logistic regression model for $A_k$ with a separate intercept for each $k$, in the subset of individuals with $A_{k-1}=0$. The probabilities $\Pr(A_k|\bar A_{k-2},A_{k-1}=0,L_0)$ were estimated using a logistic regression model for $A_k$ with $L_0$ as a covariate, and allowing both the intercept and the coefficients for $L_0$ to differ for each $k$, fitted using the subset of individuals with $A_{k-1}=0$. The probabilities $\Pr(A_k|\bar L_k,\bar A_{k-2},A_{k-1}=0)$ were estimated using a logistic regression model for $A_k$ with $L_{k}$ as a covariate in the subset of individuals with $A_{k-1}=0$, which is the correct model under our data generating mechanism. Similar models were used to estimate the weights for the sequential trials analysis, but with $L_0$ replaced by $L_k$ in trial $k$ for the model used to estimate the probabilities in the numerator of the weights in (\ref{eq:wt.seq}). The models were fitted across all trials combined, which was valid under our data generating mechanism.




\emph{Performance measures}

We assess the performance of the methods in terms of bias and efficiency of the risk difference estimates. To obtain the bias we need to know the true values. True values for the survival probabilities and the risk difference were obtained by generating data as though from a large randomized controlled trial, as explained in \cite{Keogh:2021}. For this, we first generated $L_0$ for 1 million individuals, according to the model outlined in Table 1. Two data sets are then created, one in which the 1 million individuals are set to have $A_k=1$ ($t=0,\ldots,4$) (`always treated') and another in which they are all set to have $A_t=0$ ($k=0,\ldots,4$) (`never treated'). In each data set the $L_k$ ($k=1,\ldots,4$) are then generated sequentially using the model for $L_k$ in Table 1, with $L_k$ being generated conditional on $L_{k-1}$, $U$ and $A_k$. Event times were generated in each data set according to the conditional additive hazard model in (\ref{eq:sim.condhaz}). The true survival probabilities, and corresponding risk differences, under each treatment strategy (`always treated' and `never treated') were then obtained using Kaplan-Meier estimates. 

Plots are used to present survival probability estimates and bias and efficiency results for the risk differences at time horizons $\tau$ in the range from 0 to 5. We also present results for the survival probabilities and risk differences at time horizons $\tau=1,2,3,4,5$ in tables. Estimates are accompanied by Monte Carlo errors. 

	\begin{table}
		\caption{Simulation scenarios: data generating mechanisms}
        \label{tab:sim.scen}
\begin{center}
\begin{tabular}{ll}
	\hline
    &Data generating mechanism\\
    \hline
    General data&$U\sim N(0,0.1)$\\
     generating mechanism&$L_0\sim N(U,1)$, $L_k \sim N(\delta_0+\delta_L L_{k-1}+\delta_A A_{k-1}+\delta_T k+U,1)$ ($k\geq 1$)\\
    &\\
    &$\mathrm{logit} \Pr(A_0=1|L_0)=\gamma_0+\gamma_L L_0$\\
    &$\mathrm{logit} \Pr(A_k=1|\bar A_{k-1},\bar L_{k})=\gamma_0+\gamma_A A_{k-1}+\gamma_L L_k$, ($k\geq 1$)\\
    &\\
    &$h(t|\bar A_{\lfloor t \rfloor},\bar{L}_{\lfloor t \rfloor})=\alpha_{0}+\alpha_{A}A_{\lfloor t \rfloor}+\alpha_{L}L_{\lfloor t \rfloor}+\alpha_U U$\\
    &\\
    Scenario 1&$\delta_0=0,\delta_L=0.8,\delta_A=-1,\delta_T=0.1$\\
    &$\gamma_0=-1,\gamma_L=0.5$\\
    &$\alpha_0=0.2,\alpha_A=-0.04,\alpha_L=0.015,\alpha_U=0.015$\\
        &\\
    Scenario 2&$\delta_0,\delta_L,\delta_A,\delta_T$: as in scenario 1\\
    &$\gamma_0=-3,\gamma_L=0.5$\\
    &$\alpha_0,\alpha_A,\alpha_L,\alpha_U$: as in scenario 1\\
        &\\
    Scenario 3&$\delta_0,\delta_L,\delta_A,\delta_T$: as in scenario 1\\
    &$\gamma_0=-1,\gamma_L=3$\\
    &$\alpha_0,\alpha_A,\alpha_L,\alpha_U$: as in scenario 1\\
    \hline
    \end{tabular}
    \end{center}
    \end{table}

\subsection{Results}

\begin{figure}
    \centering
        \caption{Simulation results: mean estimated survival curves obtained using the sequential trials analysis and the MSM-IPTW analysis. The faded lines show the estimates from each of the 1000 simulated data sets and the thick lines are the point-wise averages.}
    \label{fig:sim.surv}
    \includegraphics[scale=0.6]{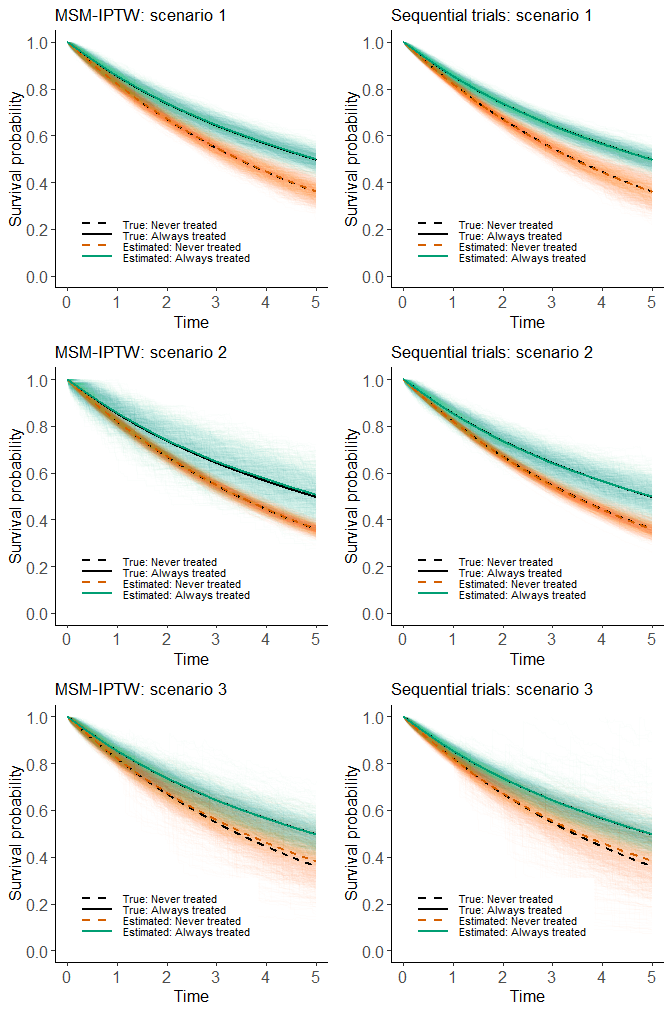}
\end{figure}

\begin{figure}
    \centering
        \caption{Simulation results: bias in estimation of the risk difference using the sequential trials analysis and the MSM-IPTW analysis. The black line shows the bias at each time point and the grey area shows the Monte-Carlo 95\% CI at each time point.}
    \label{fig:sim.bias}
    \includegraphics[scale=0.6]{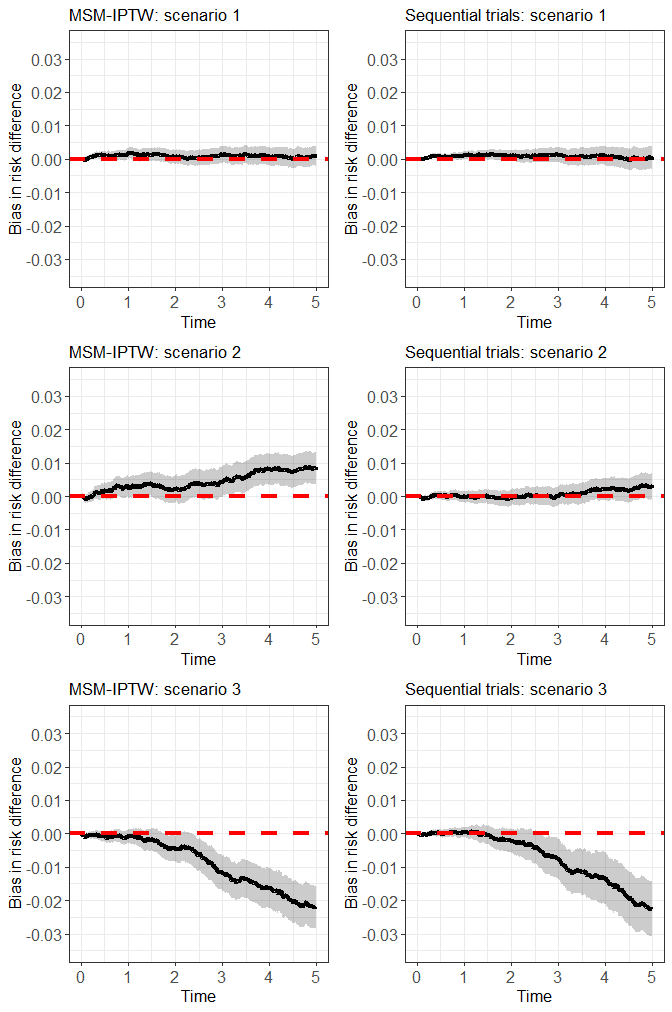}
\end{figure}

\begin{figure}
    \centering
        \caption{Simulation results: relative efficiency  of the sequential trials analysis compared with the MSM-IPTW analysis, defined as the ratio of the empirical variances of the risk difference estimates at each time.}
    \label{fig:sim.eff}
    \includegraphics[scale=0.4]{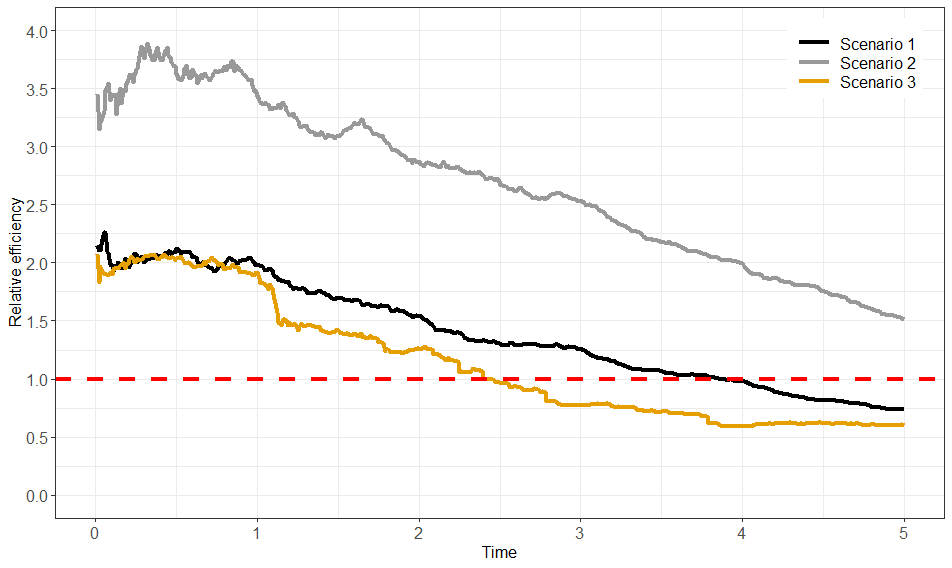}
\end{figure}

\begin{landscape}
\begin{figure}
    \centering
        \caption{Simulation results: Plots of the largest weight (by time period) in each of the 1000 simulated data sets under the sequential trials analysis (IPACW) compared with the MSM-IPTW analysis. The weights are time-dependent and change at event visit $k=0,\ldots,4$. We obtained the largest weight in each time period, where in the sequential trials the time refers to time since the start of the trial. In the sequential trials analysis, the IPACW are equal to 1 up to time 1.}
    \includegraphics[scale=0.38]{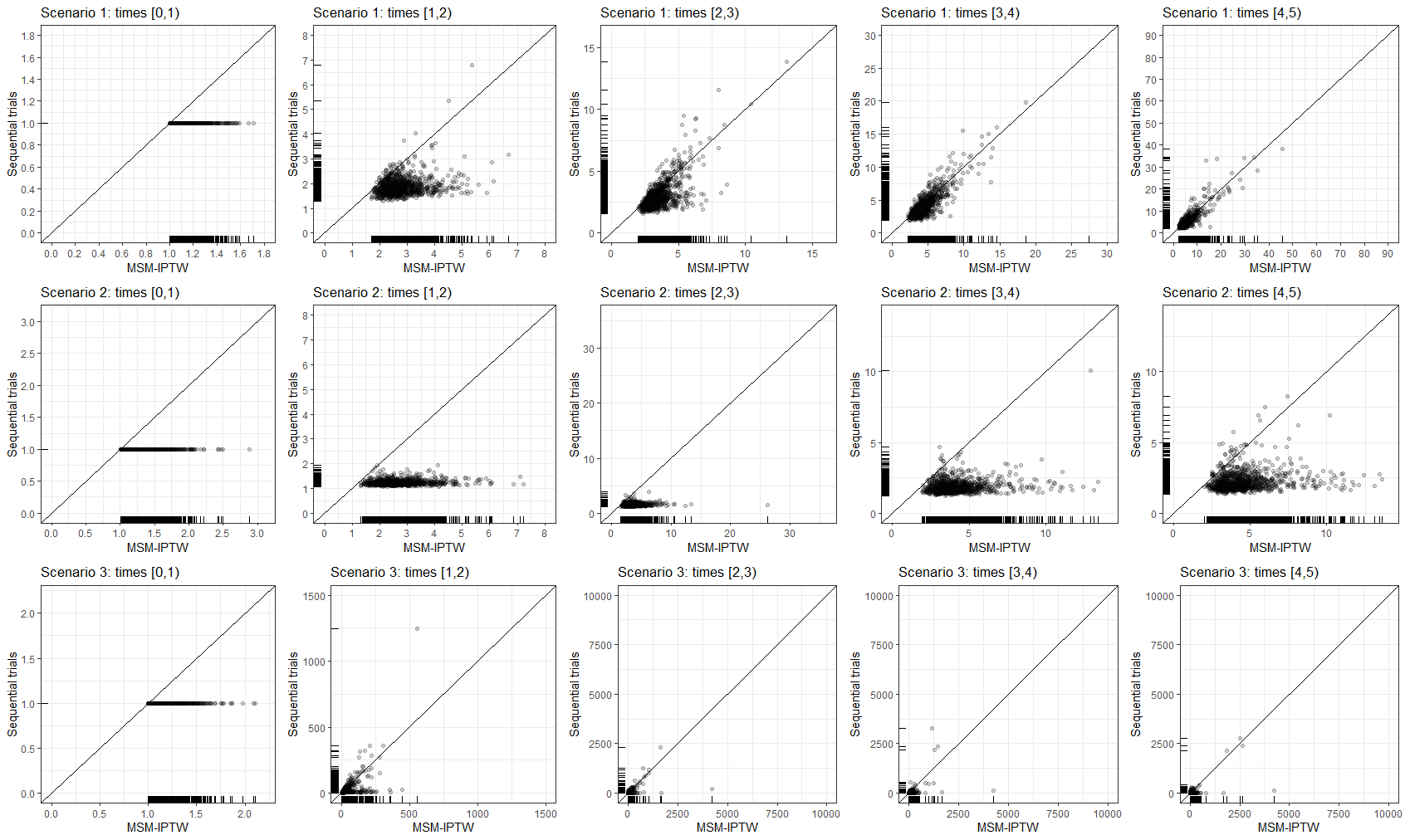}
    \label{fig:weights}
\end{figure}
\end{landscape}

For the MSM-IPTW approach we focus on the results obtained using the MSM conditional on $L_0$. Corresponding results for the MSM that is not conditional on $L_0$ are shown in the Supplementary Materials (Supplementary Figures A1 and A2). Figure \ref{fig:sim.surv} shows the estimated survivor curves under the `always treated' and `never treated' regimes under the three simulation scenarios, alongside the true curves. Figure \ref{fig:sim.bias} summarises the corresponding results for bias in risk difference estimates. The results are summarised numerically at time points $1,2,3,4,5$ in Table \ref{tab:sim.est}. The relative efficiency of the sequential trials approach compared with the MSM-IPTW approach is illustrated in Figure \ref{fig:sim.eff}. Figure \ref{fig:weights} shows plots of the largest weight used in the MSM-IPTW and sequential trials analyses in each of the 1000 simulated data sets, by time-period (because the weights are time-dependent). 

In scenario 1 both methods give unbiased estimates of the risk difference at all time points. The sequential trials approach is more efficient up to time 4, after which the MSM-IPTW approach is more efficient. The efficiency gain from the sequential trials approach is greatest at earlier time points and diminishes as time progresses. Figure \ref{fig:weights} shows that the sequential trials approach tends to use less extreme weights(IPACW) than the MSM-IPTW approach at the earlier time points, but that by time 4 after potential treatment initiation the IPTW do not appear to be more extreme than than the IPACW. The relative efficiency results are also a function of how the data are used and of modelling assumptions made. Under the assumption that treatment effects on the hazard are the same across trials (which is true under our data generating mechanism), the sequential trials approach uses information about the treatment effect on the hazard in time period $0<t<1$ from 5 trials, time period $1\leq t<2$ from 4 trials and so on. Only the first trial (starting at $k=0$) provides information about the time period $4\leq t<5$. The MSM-IPTW analysis draws information on the impact of treatment in time period $4\leq t<5$ from individuals who initiated treatment at times $0,1,2,3,4$, Supplementary Table A1 shows the number of individuals observed at times $t=0,1,2,3,4$ under the two approaches. The number of individuals observed at time $4$ is considerably greater under the MSM-IPTW approach compared with the sequential trials approach. 

In scenario 2, the sequential trials approach gives unbiased estimates of the risk difference at all time points. The MSM-IPTW approach gives unbiased estimates up to around time 3, with the risk difference being slightly biased upwards after time 3. Figure \ref{fig:sim.surv} shows that the survivor curve in the always treated group is slightly biased upwards, which we also see in Table \ref{tab:sim.est}. The bias is small in magnitude but statistically significant. Figure \ref{fig:weights} comparing the IPACW to the IPCW shos that the largest weights tend to be considerably higher in the MSM-IPTW analysis compared with the sequential trials analysis. The difference between the IPACW and IPTW is much greater in scenario 2 compared with scenario 1, with the IPACW tending to be much less extreme than the IPTW at all time points. In scenario 2 the proportion of individuals initiating treatment at a given visit is low. Supplementary Table A1 shows that in scenario 2 the absolute numbers of individuals in the `always treated group' is low across all visits. The bias for MSM-IPTW in scenario 2 is therefore attributed to the combination of large weights and finite sample bias due to small numbers of treated individuals. The efficiency plot in Figure \ref{fig:sim.eff} shows that the sequential trials analysis is more efficient than MSM-IPTW at all time points. From \ref{fig:sim.surv} and Table \ref{tab:sim.est}, we see that the variation in the estimates of the survival probabilities is much greater in MSM-IPTW for the `always treated' regime. For the `never treated' regime the variation is smaller using the sequential trials approach up to between times 2 and 3, after which the variation in the MSM-IPTW estimates is lower.

In scenario 3, both methods give bias in the risk difference after time 1 (Figure \ref{fig:sim.bias}). Figure \ref{fig:sim.surv} shows that the survival probabilities for the `always treated' regime are estimated without bias, but the survival probabilities for the `never treated' regime have some upwards bias at later time points. The magnitude of the bias is small and similar under both methods. Figure \ref{fig:sim.eff} shows that the sequential trials analysis is more efficient up to around time 2.5, and the MSM-IPTW analysis is more efficient thereafter. In this scenario the time-varying covariate $L_k$ is strongly predictive of treatment $A_k$ (log odds ratio 3). Looking at the distribution of the largest weights in Figure \ref{fig:weights} we see that after time 1 there are some very large weights under both methods. The largest weights tended to be higher in the MSM-IPTW approach. The number of individuals observed at each time point is greater under scenario 3 than scenario 2. In this case therefore, the bias may be attributed to near violations of the positivity assumption. 

Results are similar from the MSM-IPTW analysis based on the MSM without conditioning on $L_0$. We also applied the analyses using truncated weights in which both IPTW and IPACW were truncated at the 95th percentile, which resulted in similar findings - see Supplementary Materials (Supplementary Figure A3 and Figure A4). 

\section{Application in the UK CF Registry}
\label{sec:example}

\subsection{The target trial}

We apply the methods to the motivating application introduced in Section \ref{sec:motivation}. Table \ref{table:target.trial} summarises the target trial, which we emulate using longitudinal data on individuals observed in the UK CF Registry between 2008 and 2018, in addition to some data from up to three visits prior to 2008, which was used to ascertain eligibility (see below). 

Some initial data set-up steps were taken as follows. The outcome is the composite of death or transplant, so visits recorded after transplant were omitted. Data are intended to be collected at annual visits. However, there are some visits considerably closer and wider apart. We omitted visits that were less than 6 months after the previous visit \textit{and} where the previous and subsequent visit were less than 18 months apart. For individuals who do not have the event of interest (death or transplant) the censoring date was the earlier of 31 December 2018 and the date of the last visit plus 18 months. Where there was a period of more than 18 months after a given visit without the event or another visit being recorded, the individual was censored at the date of that visit plus 18 months, and any subsequent later visits excluded. The data allow for a maximum of 11 years of follow-up.

The eligibility criteria for the target trial are people with CF aged 12 or older and who have not used DNase for at least 3 years. For the emulated trial we identify individuals meeting these criteria according to UK CF Registry data recorded between 2008 and 2018. The emulated trial has some additional eligibility criteria that relate to the way in which the data are collected. An individual is defined as being eligible at a given visit if they have at least 3 preceding visits, with the patient recorded as not using DNase at each of those visits, and at least one visit in the prior 18 months. We used this (approximately) 3-year period of non-treatment use to minimise any bias arising from missing treatment data for an individual given year, or a missing annual review. Individuals who may have used DNase in the more distant past were included to increase sample size and because it was considered reasonable that the effect of re-starting DNase in this group would be similar to that for individuals who had never used DNase in the past. To be eligible at a given visit, individuals were also required to have observed measurements of forced
expiratory volume in 1 second as percentage predicted (FEV$_1$\%) and body mass index (BMI) z-score at the visit and the previous visit. The other elements of the protocol are the same for the emulated trial as for the target trial. The treatment strategies are (i) start using DNase and continue using it throughout follow-up, and (ii) do not start using DNase during follow-up. The aim is to estimate risk differences between the two treatment strategies up to 11 years of follow-up. 


There was a small amount of missing data in some of the time-dependent covariates. We used the last-observation-carried-forward to address this. For the genotype variable `missing' was treated as a separate category. 

\subsection{Analysis}

The MSM-IPTW analysis and the sequential trials analysis were both applied as outlined in Section \ref{sec:ttemulation} to estimate survival curves up to 11 years post-randomization under the two treatment regimes of `always treat' or `never treat' with DNase (Table \ref{table:target.trial}), and corresponding risk differences. For the MSM-IPTW analysis an individual's baseline visit ($k=0$) was defined as the first visit in the period 2008-2018 at which an individual meets the emulated trial eligibility criteria. For the sequential trials analysis individuals were assessed for eligibility at each visit from the baseline visit onwards, resulting in 11 possible trials.

DNase use is recorded at each annual review visit and refers to whether the patient has been prescribed DNase over the past year. We identified a number of potential confounders, which include both demographic and clinical patient characteristics (see Table \ref{table:baseline.summary}). These include two time-fixed covariates; sex and genotype class (a marker of the severity of the CF-causing mutation - low, high, not assigned, missing) \cite{McKone:2006}, and several time-dependent covariates; age (in years), lung function measured using FEV$_1$\%, BMI z-score,
chronic infection with \emph{Pseudomonas aeruginosa} or use of nebulized antibiotics, infection with
\emph{Staphylococcus aureus} or Methicillin-resistant \emph{Staphylococcus aureus} (MRSA), infection with
\emph{Burkholderia cepacia}, infection with \emph{Non-tuberculous mycobacteria} (NTM), CF-related diabetes,
pancreatic insufficiency, number of days on intravenous (IV) antibiotics at home or in hospital (categorized as 0, 1-14, 15-28, 29-42, 43+), other hospitalization (yes or no), use of other mucoactive treatments
(hypertonic saline, mannitol or acetylcysteine), use of oxygen therapy, use of CFTR modulators
(ivacaftor, lumacaftor/ivacaftor, or tezacaftor/ivacaftor). Of the time-dependent variables, FEV$_1$\% and BMI are measured on the day of the visit, whereas all of the other variables refer to information in the time since the previous visit (i.e. in the past year).

For each approach we obtained results using a Cox proportional hazards model and an Aalen's additive hazard model (with time-varying coefficients) for the MSMs. In the MSMs used for the MSM-IPTW analysis the MSM included current treatment and treatment in the three prior years, and is conditional on the time-fixed covariates and baseline measures of the time-varying covariates, except visit year. The MSM used in the sequential trials analysis included the same set of covariates as used in the MSM-IPTW analysis, i.e. the time-fixed covariates and time-varying covariates as measured at the visit at the start of each trial. The continuous covariates, age, FEV$_1$\%,and BMI z-score, were modelled using restricted cubic splines (with 3 knots) in the MSMs. For the sequential trials analysis, we performed tests of whether the coefficient(s) for treatment status differed by trial. There was no evidence of this and so we used a combined analysis across trials. In the analyses using the Cox proportional hazards model we assessed the proportional hazards assumption for the treatment variable(s). In the sequential trials analysis there was no evidence against the proportional hazards assumption for the treatment variable, whereas in the MSM-IPTW analysis there was evidence against the proportional hazards assumption in a joint test for the four treatment variables in the model.

Logistic regression models were used to estimate stabilized weights for use in each analysis method. In the MSM-IPTW approach the model used for the denominator of the weights included all time-fixed covariates and the current values of the time-dependent covariates. The model used for the numerator of the weights included the time-fixed covariates and baseline measures of the time-varying covariates. In these models, DNase status recorded at visit $k$ is assumed to depend on FEV$_1$\% and BMI z-score recorded at visit $k-1$, and on other time-dependent variables as recorded at visit $k$. 

In the simulation we assumed that once an individual initiates treatment they always continue treatment. However, in this application some individuals also stop treatment. In the sequential trials analysis we therefore estimated separate IPACW for artificial censoring due to switching from $A_k=0$ to $A_{k+1}=1$ and for switching from $A_k=1$ to $A_{k-1}=0$. The models for the weights included the same variables as included in the weights for the MSM-IPTW analysis. The continuous variables (Age, FEV$_1$\%, BMI z-score, visit year) were modelled using restricted cubic splines (with 3 knots) in the models for the weights.

For both approaches we also estimated stabilized weights for censoring due to loss-to-follow-up. These included the same variables as used in the treatment-related weights, with the difference that the censoring weights models for the probability of being censored between visits $k$ and $k+1$ included measures of FEV$_1$\% and BMI z-score made at visit $k$, rather than the measures from the previous visit. This is because the censoring indicator at visit $k$ refers to censoring before visit $k+1$, whereas the treatment indicator at visit $k$ refers to treatment status since the previous visit. 

There was missing data in some of the time-dependent covariates ($\sim$10\% missingness in covariates at the baseline visit). We used the last-observation-carried-forward to address this. For the genotype variable `missing' was treated as a separate category. 

We wished to estimate the same marginal survival probabilities, and hence risk differences, using each analysis method. Both methods include conditioning on covariates measured at baseline. We are therefore able to standardize to any appropriate population. For this analysis we standardize to the population of individuals meeting the eligibility criteria in 2018. 
Visit year was included in the weights models but not in the set of variables included in the MSM. This is because there can only be little information about long term follow-up for individuals in the later years, which results in poor convergence in some models if visit year is included in the MSMs. There are therefore differences between the population to which the MSM-IPTW analyses refer and that to which the sequential trials analyses refer, in terms of the marginal distribution of baseline years from 2008-2017. We consider this to be a minor difference that is unlikely to result in clinically important differences between the results from the two approaches.

Bootstrapping was used to obtain 95\% confidence intervals for the estimated survival curves under the two treatment regimes of `always treat' or `never treat' with DNase and the corresponding risk differences, using 1000 bootstrap samples and the percentile method. 

\subsection{Data overview}

The eligibility criteria were met by 3855 individuals at at least one visit. Among these individuals there were 338 events (266 deaths and 72 transplants). Of the 3517 individuals who did not have the event, 1780 (51\%) were administratively censored (defined as having the expected number of visits given their first year of meeting eligibility criteria, assuming one visit each year) and the remainder were censored due to loss-to-follow-up (which included individuals censored after a gap of more than 18 months without a visit). The characteristics of the individuals at the first visit at which the eligibility criteria were met are summarised in Table \ref{table:baseline.summary}. 

Table \ref{table:visits.summary} summarises the number of individuals contributing to the MSM-IPTW and sequential trials analyses and their treatment status. The eligibility criteria were first met in 2008 (the first year we considered) by 1729 individuals (45\%). In the sequential trials approach many individuals contribute to trials starting at multiple time points such that 18439 rows contribute information at the start of trials (Table \ref{table:visits.summary}(b)). At visit $k=10$, there remain 643 individuals in the MSM-IPTW analysis but only 241 rows in the sequential trials analysis, due to the artificial censoring in the sequential trials approach. At the first visit at which eligibility criteria were met, 612 initiated DNase treatment (16\%). Of those who initiated DNase at this first eligible visit, 69\% used DNase at all visits, and of those who did not initiate DNase at the first eligible visit 53\% were non-users at all visits. In the MSM-IPTW analysis the proportion of treated individuals is similar across visits, ranging from 15-17\%. In the sequential trials analysis the proportion of treated individuals at a given visit is equivalent to the proportion 'always treated' up to that visit (due to the artificial censoring), and this percentage increases over time from 15\% at visit $k=0$ to 22\% at visit $k=10$.

\subsection{Results}

\begin{figure}
    \centering
        \caption{Estimated survival curves in the `always' treated and `never treated' with DNase from the MSM-IPTW and sequential trials analyses. The shaded areas show the 95\% confidence intervals.}
    \label{fig:cf.surv}
    \includegraphics[scale=0.45]{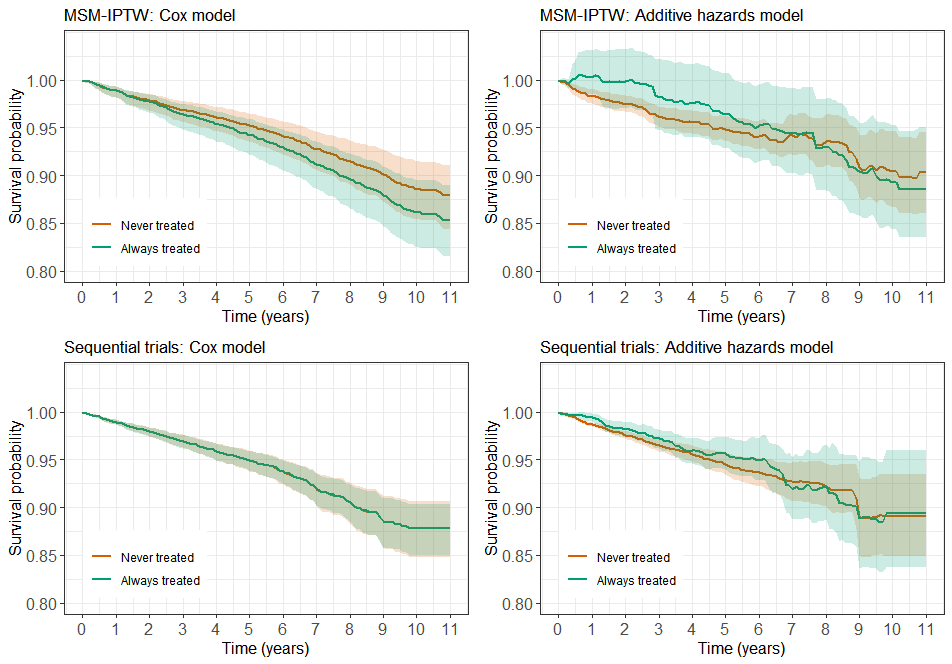}
\end{figure}

\begin{figure}
    \centering
        \caption{Estimated risk difference (`always' treated vs `never treated' with DNase) from the MSM-IPTW and sequential trials analyses. The shaded area shows the 95\% confidence intervals.}
    \label{fig:cf.riskdiff}
    \includegraphics[scale=0.45]{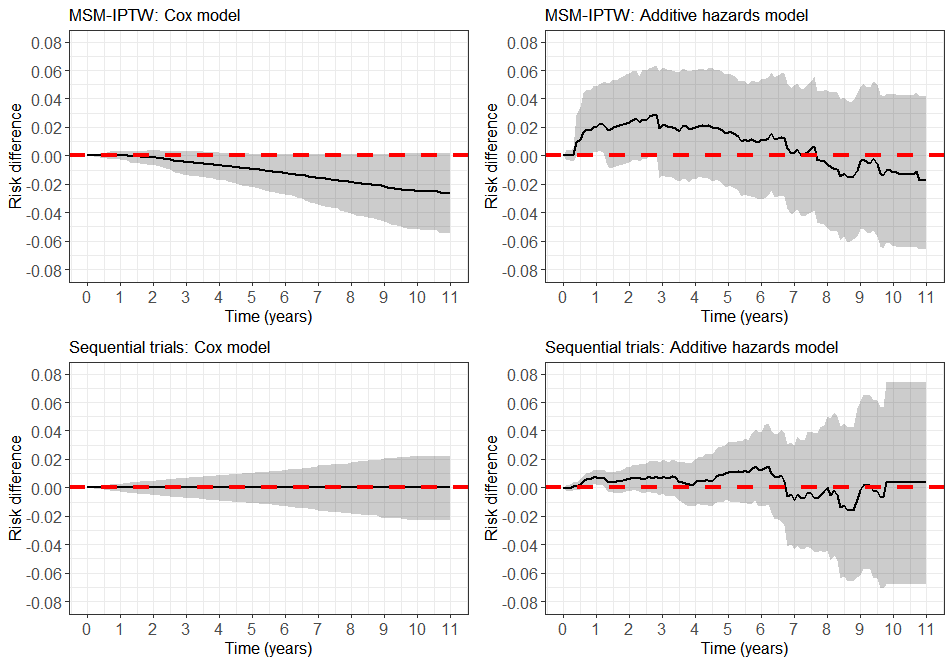}
\end{figure}

Figure \ref{fig:cf.surv} shows the estimated survivor curves from the MSM-IPTW and sequential trials analyses, using Cox proportional hazards models and additive hazards models for the MSMs. The corresponding estimated risk differences are shown in Figure \ref{fig:cf.riskdiff}, where a risk difference greater than 1 indicates better survival in the `always treated'. 

None of the results provide evidence that DNase use affects the outcome, either in the short or long term. In the MSM-IPTW analysis using a Cox proportional hazards model, the survival curve for the `always treated' regime is consistently below that for the `never treated' regime, with the differences increasing over time. In the sequential trials analysis using the Cox model the survival curves in the two treatment groups are very similar over time. Using the additive hazards model the survival curve for the `always treated' regime is above that for the `never treated' regime for the first few years, more so in the MSM-IPTW results, before the curves cross. As we would expect, the 95\% CIs from using the Cox model are narrower than those from the additive hazards model. The 95\% CIs are generally narrower under the sequential trials approach, except when using the additive hazards models in the later part of follow-up. Supplementary Figure A5 shows a plot of the distribution of the weights used under the two analysis approaches. This shows that a large number of rows are assigned a weight of 1 in the sequential trials analysis. This is because in the sequential trials data the first row corresponding to each trial has a weight of 1. There are no extreme weights in the MSM-IPTW analysis, but the distribution is less highly concentrated around 1.

\section{Discussion}
\label{sec:discussion}

In this paper we have focused on estimating the effects of longitudinal treatment regimes on survival, using observational data subject to time-varying confounding of the treatment-outcome association. We have compared the MSM-IPTW approach with the sequential trials approach. A key contribution has been to specify causal estimands that can be identified using the sequential trials approach, and we showed how the previously described implementations can be shown to estimate underlying MSMs for counterfactual outcomes under certain assumptions. The methods were considered in the context of emulating a target trial and we outlined a general protocol for a target trial for longitudinal treatment regimes and time-to-event outcomes. We compared the MSM-IPTW and sequential trials approaches in terms of how the data are used, the form of the MSMs, how time-dependent confounding is addressed through the analysis, and when they identify the same parameters. 

For time-to-event outcomes, estimands based on contrasts between risks rather than contrasts between hazards have been shown to have better causal interpretations. We outlined how the outputs from the sequential trials analysis and the MSM-IPTW analysis can be used to obtain marginal absolute risk estimates using empirical standardization, and hence to obtain marginal risk differences or ratios. Care should be taken over defining the population to which marginal risk estimates refer.  Applications of the sequential trials approach in the literature have tended to focus on presentation of hazard ratio estimates. As well as giving hazard ratio estimates, Hernan et al. (2008)\cite{Hernan:2008} presented estimated survivor curves under the treated and untreated regimes. However, it was not clear how the survival curves were obtained or what population they referred to. Cole et al. (2017)\cite{Cole:2017} and Garcia-Albinez et al (2017)\cite{Garcia:2017} also obtained survival curves using the sequential trials approach but similarly the methods were not described in detail. In this paper, we have provided clear guidance on how to obtain estimated survival curves from the MSM-IPTW and sequential trials approaches.

A key difference between the MSM-IPTW approach and the sequential trials approach is that the MSM-IPTW approach enables estimation of risks under any longitudinal treatment regime, whereas the sequential trials approach focuses only on `always treated' and `never treated' regimes. The sequential trials approach could be extended to consider different longitudinal treatment regimes, with individuals being censored when they deviate from a specified regime. This is an area for future investigation. We focused in this paper on a setting in which the trials used in the sequential trials approach have different lengths of follow-up, with trials starting at later times having shorter follow-up. It would also be possible to focus only on trials with the same fixed length of follow-up. This depends on the total length of follow-up available in the data, and on the time horizons of interest for the risk differences being estimated. For example, in the simulation we consider a situation with 5 visit times. Had we been interested in a risk difference for a time horizon of $\tau=2$, we could have used sequential trials of equal length starting at times 0,1,2, and 3. For the MSM analysis, the question would arise as to whether to focus on an analysis starting from time 0, or starting from time 3, say. Benefits of using sequential trials of equal length of follow-up could be that it is possible that each trial can estimate a separate baseline hazard covering the full length of follow-up needed for the risk difference.

Van der Laan et al. (2005)\cite{vanderLaan:2005} and Petersen and van der Laan (2007)\cite{Petersen:2007a} introduced history-adjusted MSMs (HA-MSM), which are an extension of MSMs.  In a HA-MSM, an MSM is used from a series of new time points (visits) in a study, with the MSM being conditional on covariates and treatment history up to that time, and assuming a common MSM across time points. Estimation of the MSMs is using IPTW. As well as being an extension of the MSM-IPTW approach, there is a clear connection between HA-MSMs and the sequential trials approach. Hernan and Robins (2016)\cite{Hernan:2016AJE} referenced HA-MSMs in their paper describing the target trial framework when noting that eligibility criteria could be applied at multiple time points. A distinction between HA-MSMs and the sequential trials approach is that the sequential trials approach restricts to individuals who have not yet started the treatment at each new time origin, whereas the HA-MSM approach conditions on past treatment in the MSM. Like in MSMs, the HA-MSM approach considers all longitudinal treatment regimes, in contrast to the sequential trials approach, which focuses only on two treatment regimes. An example of the use of HA-MSMs in the context of a survival outcome was given by Bembom et al.(2009)\cite{Bembom:2009}. However, the HA-MSM approach does not yet appear to have been much used in practice, perhaps due to initial criticisms of the method \citep{Robins:2007,Petersen:2007b}. The sequential trials approach is also related to the sequential stratification method described by Schaubel et al. (2006)\cite{Schaubel:2006}, in which individuals initiating treatment at a given time point are matched to untreated individuals in a sequential manner over time. The analysis results in an estimate of the treatment effect on the treated. An area for future work is to further discuss the different sequential methods for causal inference for survival outcomes, in particularly to compare the estimands that they focus on and their assumptions.

In this paper we compared the MSM-IPTW and sequential trials approaches using a simulation study, focusing on bias and efficiency. We found that the sequential trials approach can be more efficient than the MSM-IPTW approach for estimating risk differences, but that this is not true at all time points: for risks at time points corresponding to longer follow-up the MSM-IPTW approach can be more efficient. We linked the efficiency gains from the sequential trials approach to the use of less extreme weights (IPACW) compared with the IPTW. It must also be recognised that efficiency gains from the sequential trials approach at some time points arises in part due to assumptions about consistency of the treatment effect across different trials (i.e. from different time origins). Throughout the paper we discussed the methods in terms of Cox proportional hazard models or Aalen additive hazard models for the MSM. Previous work in this context has focused on Cox models for the MSM. We have shown that the additive hazard model can have some advantages in terms of relaxing assumptions - for example, it naturally allows the hazard to depend on treatment duration in the sequential trials approach. Use of the the additive hazard model was also advantageous in the simulation study as it ensured that the analysis models used in both the MSM-IPTW approach and in the sequential trials approach could both be correctly specified, enabling a fair comparison of the methods. There have been few simulation comparisons of causal methods for time-to-event outcomes due to the challenges of  making fair comparisons. As noted earlier, Karim et al.(2018)\cite{Karim:2018} compared the MSM-IPTW and sequential trials approaches but focused on estimation of hazard ratios and compared efficiency of marginal and conditional hazard ratios, which are not the same quantity. In further work it would be of interest to use a similar simulation approach to compare other methods, including the other sequential methods noted above. 

The methods were applied to investigate the effect of the treatment DNase for people with CF on the composite outcome of death or transplant using longitudinal observational data from the UK CF Registry. We did not find any evidence of a causal treatment effect using either the MSM-IPTW approach or the sequential trials approach. In recent work, longitudinal data from the UK CF Registry was used to estimate the longer term causal effect of DNase on the non-time-to-event outcomes of lung function and number of days of intravenous antibiotic medication using MSM-IPTW, g-computation and g-estimation \cite{Newsome:2018SIM,Newsome:2019JCF}. This work found evidence of a benefit of DNase for people with low lung function, but not for those whose lung function was higher. Seaman et al. (2020)\cite{Seaman:2020} illustrated the structural nested cumulative survival time approach with an investigation of the effect of DNase use on survival using UK CF Registry data, and also did not find evidence of a treatment effect. Sawicki et al (2012)\cite{Sawicki:2012} used longitudinal data from the US Cystic Fibrosis Foundation patient Registry to investigate the effect of DNase treatment on the risk of death in the next year using pooled logistic regression analyses, finding evidence of a reduced risk of death with DNase use. Limitations of our analysis include that Dnase is used by a large proportion of the population from a young age, and we focused on individuals initiating DNase aged 12 and older, as there are few deaths in younger individuals. It is possible that patients initiating DNase at older ages have particular characteristics that are also associated with survival, and which we did not fully capture in the set of adjustment variables used in our analysis. The number of events in our data was relatively low, which at least in part explains the wide uncertainty on our estimates. Our analysis included a relatively large number of adjustment variables compared to the number of events, which included several continuous variables, and we did not include any interaction terms in the weights models or the MSMs. In further work it would be of interest to investigate data-adaptive methods for flexible adjustment for the time-dependent confounders. Another important consideration in interpreting our results is that the UK CF Registry records DNase prescription, but adherence to the treatment is unknown and may be low and dependent on individual characteristics. In further work it would be of interest to use Registry data to investigate the impact of removing DNase treatment, particularly in light of new disease-modifying treatments that have been introduced for CF in recent years. 

In summary, the MSM-IPTW and sequential trials approaches can both be used to estimate causal effects of time-varying treatments using observational data, controlling for time-dependent confounding. The sequential trials approach tends to use less extreme weights, which can result in substantial gains in efficiency relative to the MSM-IPTW approach. Both approaches rely on modelling assumptions. The MSM used in MSM-IPTW is prone to misspecification as it considers all treatment patterns, whereas the sequential trials approach focuses only on two regimes. However, the sequential trials approach is prone to model misspecification when a combined analysis across trials is used, and this risk of misspecification is arguably greater than that in the MSM-IPTW approach. The sequential trials approach is appealing in that it makes efficient use of longitudinal data that are often available in observational databases, and that it explicitly mimics a hypothetical trial. There are several areas for further development of this approach that could extend the types questions that it could be used to address.

\vspace{1cm}

{\bf Funding}

RHK is funded by a UK Research \& Innovation Future Leaders Fellowship (MR/S017968/1), JMG by the Research Council of Norway (Grant No. 273674), SRS by MRC programme grant MC\_UU\_00002/10, and GD by a UK Research \& Innovation Future Leaders Fellowship (MR/T041285/1).

{\bf Data availability}

This work used anonymized data from the UK Cystic Fibrosis Registry, which has Research Ethics Approval (REC ref: 07/Q0104/2). The use of the data was approved by the Registry Research Committee. Data are available following application to the Registry Research Committee\\ (https://www.cysticfibrosis.org.uk/the-work-we-do/uk-cf-registry/apply-for-data-from-the-uk-cf-registry). The authors thank people with CF and their families for consenting to their data being held in the UK CF Registry, and NHS teams in CF centres and clinics for the input of data into the Registry. 

\begin{table}[]
    \centering
        \caption{Simulation results: Estimated survival probabilities ('Est') under the always treated (S1) and never treated (S0) regimes and risk difference(RD) (mean over 1000 simulations) at time $\tau$, with corresponding empirical standard deviation (over 1000 simulations). Corresponding bias in the estimates ($\times 10^4$), accompanied by the MC error ($\times 100$).}
    \label{tab:sim.est}
    \begin{tabular}{llrrrrr}
    \hline
    &&\multicolumn{5}{c}{Time $\tau$}\\
    &&1&2&3&4&5\\
    \hline
    \multicolumn{7}{l}{MSM-IPTW: scenario 1}\\
S1 & Est (SD) & 0.854 (0.021) & 0.738 (0.027) & 0.644 (0.030) & 0.566 (0.031) & 0.498 (0.031) \\ 
   & Bias (MC error) & 0.170 (0.066) & 0.166 (0.084) & 0.229 (0.094) & 0.223 (0.097) & 0.189 (0.099) \\ 
  S0 & Est (SD) & 0.819 (0.015) & 0.671 (0.019) & 0.547 (0.024) & 0.445 (0.029) & 0.361 (0.033) \\ 
   & Bias (MC error) & 0.002 (0.047) & 0.114 (0.061) & 0.100 (0.076) & 0.137 (0.091) & 0.098 (0.105) \\ 
  RD & Est (SD) & 0.034 (0.026) & 0.067 (0.033) & 0.097 (0.040) & 0.121 (0.044) & 0.138 (0.047) \\ 
   & Bias (MC error) & 0.168 (0.082) & 0.053 (0.106) & 0.129 (0.126) & 0.086 (0.139) & 0.092 (0.148) \\
 \addlinespace
     \multicolumn{7}{l}{Sequential trials: scenario 1}\\
S1 & Est (SD) & 0.853 (0.015) & 0.736 (0.019) & 0.641 (0.022) & 0.563 (0.024) & 0.496 (0.027) \\ 
   & Bias (MC error) & 0.068 (0.047) & 0.010 (0.059) & -0.068 (0.068) & -0.086 (0.076) & -0.037 (0.085) \\ 
  S0 & Est (SD) & 0.819 (0.010) & 0.669 (0.017) & 0.544 (0.025) & 0.442 (0.033) & 0.359 (0.044) \\ 
   & Bias (MC error) & -0.043 (0.031) & -0.085 (0.054) & -0.202 (0.079) & -0.191 (0.105) & -0.076 (0.140) \\ 
  RD & Est (SD) & 0.034 (0.018) & 0.067 (0.027) & 0.097 (0.036) & 0.121 (0.044) & 0.137 (0.055) \\ 
   & Bias (MC error) & 0.111 (0.058) & 0.094 (0.085) & 0.134 (0.112) & 0.105 (0.140) & 0.039 (0.173) \\ 
  \addlinespace
      \multicolumn{7}{l}{MSM-IPTW: scenario 2}\\
S1 & Est (SD) & 0.855 (0.049) & 0.739 (0.063) & 0.648 (0.071) & 0.573 (0.072) & 0.506 (0.074) \\ 
   & Bias (MC error) & 0.333 (0.154) & 0.343 (0.199) & 0.569 (0.224) & 0.934 (0.227) & 0.933 (0.233) \\ 
  S0 & Est (SD) & 0.819 (0.012) & 0.671 (0.015) & 0.547 (0.016) & 0.445 (0.018) & 0.361 (0.018) \\ 
   & Bias (MC error) & 0.018 (0.039) & 0.123 (0.048) & 0.133 (0.052) & 0.112 (0.055) & 0.077 (0.056) \\ 
  RD & Est (SD) & 0.036 (0.050) & 0.069 (0.065) & 0.100 (0.073) & 0.128 (0.075) & 0.145 (0.077) \\ 
   & Bias (MC error) & 0.315 (0.159) & 0.219 (0.206) & 0.436 (0.231) & 0.823 (0.237) & 0.856 (0.243) \\ 
   \addlinespace
      \multicolumn{7}{l}{Sequential trials: scenario 2}\\
S1 & Est (SD) & 0.852 (0.026) & 0.735 (0.036) & 0.640 (0.042) & 0.564 (0.048) & 0.498 (0.056) \\ 
   & Bias (MC error) & -0.032 (0.083) & -0.125 (0.112) & -0.162 (0.132) & 0.038 (0.150) & 0.195 (0.178) \\ 
  S0 & Est (SD) & 0.819 (0.007) & 0.669 (0.012) & 0.544 (0.017) & 0.442 (0.020) & 0.359 (0.024) \\ 
   & Bias (MC error) & -0.030 (0.023) & -0.088 (0.039) & -0.184 (0.052) & -0.189 (0.063) & -0.128 (0.075) \\ 
  RD & Est (SD) & 0.033 (0.027) & 0.066 (0.038) & 0.096 (0.046) & 0.122 (0.053) & 0.140 (0.062) \\ 
   & Bias (MC error) & -0.001 (0.086) & -0.036 (0.122) & 0.022 (0.145) & 0.227 (0.168) & 0.324 (0.197) \\ 
   \addlinespace
         \multicolumn{7}{l}{MSM-IPTW: scenario 3}\\
S1 & Est (SD) & 0.852 (0.023) & 0.736 (0.034) & 0.642 (0.042) & 0.564 (0.046) & 0.497 (0.049) \\ 
   & Bias (MC error) & 0.013 (0.074) & 0.044 (0.108) & -0.007 (0.134) & 0.005 (0.146) & 0.001 (0.154) \\ 
  S0 & Est (SD) & 0.820 (0.018) & 0.674 (0.040) & 0.558 (0.056) & 0.460 (0.067) & 0.382 (0.072) \\ 
   & Bias (MC error) & 0.100 (0.056) & 0.478 (0.125) & 1.165 (0.178) & 1.631 (0.213) & 2.181 (0.229) \\ 
  RD & Est (SD) & 0.032 (0.033) & 0.062 (0.065) & 0.084 (0.086) & 0.104 (0.097) & 0.115 (0.103) \\ 
   & Bias (MC error) & -0.087 (0.106) & -0.434 (0.204) & -1.172 (0.271) & -1.626 (0.307) & -2.180 (0.327) \\ 
       \multicolumn{7}{l}{Sequential trials: scenario 3}\\
S1 & Est (SD) & 0.853 (0.017) & 0.736 (0.027) & 0.642 (0.044) & 0.564 (0.060) & 0.497 (0.060) \\ 
   & Bias (MC error) & 0.046 (0.053) & 0.005 (0.087) & 0.021 (0.138) & 0.079 (0.189) & 0.077 (0.191) \\ 
  S0 & Est (SD) & 0.819 (0.013) & 0.672 (0.036) & 0.554 (0.063) & 0.458 (0.080) & 0.383 (0.091) \\ 
   & Bias (MC error) & -0.005 (0.041) & 0.215 (0.115) & 0.775 (0.198) & 1.427 (0.254) & 2.344 (0.287) \\ 
  RD & Est (SD) & 0.033 (0.024) & 0.064 (0.058) & 0.088 (0.098) & 0.106 (0.126) & 0.114 (0.133) \\ 
   & Bias (MC error) & 0.051 (0.076) & -0.211 (0.182) & -0.754 (0.309) & -1.347 (0.399) & -2.267 (0.420) \\ 
 \hline
    \end{tabular}
\end{table}

\begin{table}
    \centering
	\caption{Summary of the protocol of a target trial for studying the effect of DNase use on survival in people with cystic fibrosis (CF).}
	\label{table:target.trial}
	\begin{tabular}{lp{10cm}}
		\hline
		Protocol component&\\
		\hline
		Eligibility criteria&People with CF between 2008 and 2017 and aged 12 or older and who have not used DNase for at least 3 years.\\
		&\\
		Treatment strategies&(i) Do not start using DNase during follow-up. (ii)
		Start using DNase and continue to use it throughout follow-up. 
		\\
		&\\
		Assignment procedures&Patients randomly assigned to either treatment strategy. Patients and their health care teams are aware of the patient's treatment status. 
		\\
		&\\
		Follow-up period&Starts at randomization and ends at death or transplant, loss-to follow-up, or the end of 2018, whichever occurs first. \\
		&\\
		Outcome&Death or transplant.\\
		&\\
		Causal contrasts of interest&Per-protocol treatment effect. Survival curves up to 11 years under the two treatment strategies, and corresponding risk differences.\\
		\hline
	\end{tabular}
\end{table}

\begin{table}[ht]
	\caption{Summary of individual characteristics at the first visit at which eligibility criteria were met.}\label{table:baseline.summary}
	\centering
\begin{tabular}{p{10cm}ll}
  \hline
 Variable 		&Mean (SD)& Median (IQR)\\
\hline
Age (years) & 24.32 (11.56) & 21.6 (14.39,30.54) \\ 
FEV$_1$\% & 76.12 (21.93) & 78.8 (62.2,92.38) \\ 
BMI z-score & 0.15 (1.22) & 0.14 (-0.6,0.96) \\ 
 \addlinespace
 		&N& \%\\
		\cline{2-3}
{\bf Sex} &&\\
Male & 2120 & (55\%)\\ 
Female &  1735 & (45\%) \\ 
   \addlinespace
{\bf Genotype class}&&\\
High & 2586 & (67\%) \\  
Low & 596 & (15\%) \\ 
Not assigned & 594 & (15\%) \\ 
Missing & 79 & (2\%) \\ 
 \addlinespace
{\bf Presence of airway infections}&&\\
\emph{Staphylococcus aureus} or Methicillin-resistant \emph{Staphylococcus aureus} (MRSA) & 1676 & (43\%) \\ 
\emph{Pseudomonas Aeruginosa} (or use of nebulized antibiotics) & 2453 & (64\%) \\ 
\emph{Burkholderia cepacia} & 113 & (3\%) \\ 
\emph{Non-tuberculous mycobacteria} (NTM) & 162 & (4\%) \\ 
  \addlinespace 
{\bf Presence of complications}&&\\
CF-related diabetes& 707 & (18\%) \\ 
Pancreatic insufficiency& 3028 & (79\%) \\ 
    \addlinespace 
{\bf Other treatments}&&\\
Use of other mucoactive treatments (hypertonic saline, mannitol or acetylcysteine)& 554 & (14\%) \\ 
Use of CFTR modulators
(ivacaftor, lumacaftor/ivacaftor, or tezacaftor/ivacaftor)& 54 & (1\%) \\ 
Use of oxygen therapy& 153 & (4\%) \\ 
Hospitalisation (not for IV antibiotics)& 145 & (4\%) \\
    \addlinespace 
Days using IV antibiotics& &  \\ 
 0 & 2315 & (60\%) \\ 
1-14 & 640 & (17\%) \\ 
  15-28 & 329 & (9\%) \\ 
  29-42 & 227 & (6\%) \\ 
  $>$42 & 344 & (9\%) \\ 
   \hline
\end{tabular}
\end{table}

\begin{table}[ht]
\caption{Summary of number of individuals contributing to the MSM-IPTW and sequential trials analyses and numbers untreated and treated: (a) by year, (b) by visit number. }
\label{table:visits.summary}

(a) By year
\begin{center}
\begin{tabular}{rlllllll}
  \hline
  &\multicolumn{3}{c}{MSM-IPTW}&&\multicolumn{3}{c}{Sequential trials}\\
Year & Number observed & Untreated & Treated & & Number observed & Untreated & Treated\\ 
\cline{2-4}\cline{6-8}
2008 & 1729 & 1452 (84\%) & 277 (16\%) & &1729 & 1452 (84\%) & 277 (16\%) \\ 
  2009 & 537 & 440 (82\%) & 97 (18\%) & &1908 & 1599 (84\%) & 309 (16\%) \\ 
  2010 & 352 & 285 (81\%) & 67 (19\%) && 2004 & 1637 (82\%) & 367 (18\%) \\ 
  2011 & 359 & 310 (86\%) & 49 (14\%) & &2039 & 1731 (85\%) & 308 (15\%) \\ 
  2012 & 264 & 234 (89\%) & 30 (11\%) & &2040 & 1728 (85\%) & 312 (15\%) \\ 
  2013 & 179 & 156 (87\%) & 23 (13\%) & &1916 & 1652 (86\%) & 264 (14\%) \\ 
  2014 & 152 & 127 (84\%) & 25 (16\%) && 1866 & 1601 (86\%) & 265 (14\%) \\ 
  2015 & 123 & 100 (81\%) & 23 (19\%) && 1779 & 1519 (85\%) & 260 (15\%) \\ 
  2016 & 90 & 79 (88\%) & 11 (12\%) & &1629 & 1426 (88\%) & 203 (12\%) \\ 
  2017 & 70 & 60 (86\%) & 10 (14\%) & &1529 & 1325 (87\%) & 204 (13\%) \\ 
   \hline
\end{tabular}
\end{center}
\vspace{0.5cm}

(b) By visit number
\begin{center}
\begin{tabular}{rlllllll}
  \hline
  &\multicolumn{3}{c}{MSM-IPTW}&&\multicolumn{3}{c}{Sequential trials}\\
Visit & Number observed & Untreated & Treated & & Number observed & Untreated & Treated\\ 
\cline{2-4}\cline{6-8}
0 & 3855 & 3243 (84\%) & 612 (16\%) && 18439 & 15670 (85\%) & 2769 (15\%) \\ 
1 & 3466 & 2901 (84\%) & 565 (16\%) && 16889 & 14321 (85\%) & 2568 (15\%) \\ 
2& 3121 & 2609 (84\%) & 512 (16\%) && 12302 & 10347 (84\%) & 1955 (16\%) \\ 
3 & 2821 & 2350 (83\%) & 471 (17\%) && 8958 & 7408 (83\%) & 1550 (17\%) \\ 
4 & 2500 & 2101 (84\%) & 399 (16\%) && 6308 & 5196 (82\%) & 1112 (18\%) \\ 
5 & 2205 & 1851 (84\%) & 354 (16\%) && 4380 & 3564 (81\%) & 816 (19\%) \\ 
6 & 1870 & 1565 (84\%) & 305 (16\%)& & 2929 & 2335 (80\%) & 594 (20\%) \\ 
7 & 1570 & 1308 (83\%) & 262 (17\%) && 1899 & 1490 (78\%) & 409 (22\%) \\ 
8 & 1246 & 1036 (83\%) & 210 (17\%) && 1154 & 882 (76\%) & 272 (24\%) \\ 
9 & 972 & 824 (85\%) & 148 (15\%) && 609 & 472 (78\%) & 137 (22\%) \\ 
10 & 643 & 548 (85\%) & 95 (15\%) && 241 & 187 (78\%) & 54 (22\%) \\ 
   \hline
\end{tabular}
\end{center}
\end{table}

\clearpage

\bibliographystyle{WileyNJD-AMA}
\bibliography{references}

\pagebreak

\include{seq_trials_supp_arxiv}
\end{document}

%% file: seq_trials_supp_arxiv.tex
\setcounter{section}{0}
\setcounter{equation}{0}
\setcounter{figure}{0}
\setcounter{table}{0}
\setcounter{page}{1}

\renewcommand{\thesection}{A\arabic{section}}
\renewcommand{\theequation}{A\arabic{equation}}
\renewcommand{\thefigure}{A\arabic{figure}}
\renewcommand{\thetable}{A\arabic{table}}

\makeatletter
\makeatother

\begin{center}
\Large{Causal inference in survival analysis using longitudinal observational data: Sequential trials and marginal structural models}

\Large{Supplementary Material}
\end{center}

\section{Inverse probability weights}

\subsection{MSM-IPTW}

To estimate MSMs using IPTW, the weight at time $t$ for individual $i$ is the inverse of their probability of their observed treatment pattern up time time $t$ given their time-dependent covariate history \citep{Daniel:2013,Cole:2008}:
\begin{equation}
W_i(t)=\prod_{k=0}^{\lfloor t \rfloor}\frac{1}{\Pr(A_k=A_{k,i}|\bar{L}_{k,i},\bar{A}_{k-1,i},T\geq k)}
\end{equation}
Some individuals can have very large weights, which can results in the parameters of the MSM being estimated very imprecisely, and therefore stabilized weights are typically used. The stabilized weight for individual $i$ is:
\begin{equation}
SW_i(t)=\prod_{k=0}^{\lfloor t \rfloor}\frac{\Pr(A_k=A_{k,i}|\bar{A}_{k-1,i},T\geq k)}{\Pr(A_k=A_{k,i}|\bar{L}_{k,i},\bar{A}_{k-1,i},T\geq k)}
\label{eq:stab.wt1}
\end{equation}
The model in the numerator of the stabilized weights can also include covariates measured at time 0, $\bar L_0$:
\begin{equation}
SW_i(t)=\prod_{k=0}^{\lfloor t \rfloor}\frac{\Pr(A_k=A_{k,i}|\bar{A}_{k-1,i},\bar L_{0,i},T\geq k)}{\Pr(A_k=A_{k,i}|\bar{L}_{k,i},\bar{A}_{k-1,i},T\geq k)},
\end{equation}
in which case the MSM must also condition on these variables, as noted in the main text.

The models used in the weights can be estimated using logistic regression, and they may be fitted using all visits combined or separately by visit. 

\subsection{Sequential trials}

For individuals in the trial starting at visit $k$ but who do not initiate treatment at that time ($\bar A_k=0$) the weight at time $t-k$ from the start of the trial is
\begin{equation}
W^0_{i,k}(t-k)=\prod_{j=k+1}^{k+\lfloor t \rfloor }\frac{1}{\Pr(A_{j}=0|\bar{L}_{j,i},\bar{A}_{(k,j-1),i}=0,T\geq k)}
\end{equation}
where $\bar{A}_{(k,j-1)}=\{A_k,A_{k+1},\ldots,A_{j-1}\}$ denotes the treatment status from visit $k$ to visit $j-1$. For those who initiate treatment at the start of trial $k$ ($\bar A_{k-1}=0,A_k=1$) the weight is 
\begin{equation}
W^1_{i,k}(t-k)=\prod_{j=k+1}^{k+\lfloor t \rfloor }\frac{1}{\Pr(A_{j}=1|\bar{L}_{j,i},\bar{A}_{k-1,i}=0,\bar{A}_{(k,j-1),i}=1,T\geq k)}
\end{equation}
In the trial starting at visit $k$, the weights are equal to 1 for all individuals between time $k$ and $k+1$, i.e. up to follow-up time 1 in trial $k$. 

The corresponding stabilized weights are
\begin{equation}
SW^0_{i,k}(t-k)=\prod_{j=k+1}^{k+\lfloor t \rfloor}\frac{\Pr(A_{j}=0|\bar{L}_{k,i},\bar{A}_{(k,j-1),i}=0,T\geq k)}{\Pr(A_{j}=0|\bar{L}_{j,i},\bar{A}_{(k,j-1),i}=0,T\geq k)}
\end{equation}
and
\begin{equation}
SW^1_{i,k}(t-k)=\prod_{j=k+1}^{k+\lfloor t \rfloor }\frac{\Pr(A_{j}=1|\bar{L}_{k,i},\bar{A}_{k-1,i}=0,\bar{A}_{(k,j-1),i}=1,T\geq k)}{\Pr(A_{j}=1|\bar{L}_{j,i},\bar{A}_{k-1,i}=0,\bar{A}_{(k,j-1),i}=1,T\geq k)}
\end{equation}
where the probability in the numerator is conditional on the covariates $\bar L_k$ observed at the start of trial $k$. 

The models used in the weights can be estimated using logistic regression, and they may be fitted using all visits combined or separately by visit, and across all trials combined or separately by trial.

\section{Assumptions}

The no interference assumption is that the counterfactual event time for a given individual, $T^{\underline{a}_{0}}$, does not depend on the treatment received by any other individuals. The positivity assumption is that each individual has a strictly non-zero probability of receiving each given pattern of treatments over time. Consistency means that an individual's observed outcome is equal to the counterfactual outcome when the assigned treatment pattern is set to that which was actually received, $T_i=T_i^{\underline{A}_{0,i}}$. The conditional exchangeability assumption can be expressed formally as $T^{\bar{A}_{k-1},\underline{a}_{k}}\indep A_k|\bar{A}_{k-1},\bar{L}_{k},T\geq k$ for all feasible $\underline{a}_{k}$, where $T^{\bar{A}_{k-1},\underline{a}_{k}}$ denotes the counterfactual event time had an individual followed their observed treatment pattern up to time $k-1$, $\bar{A}_{k-1}$, and had their treatments been set to $\underline{a}_{k}$ from time $k$ onwards, given survival to time $k$. The conditional exchangeability assumption means that among individuals who remain at risk of the event at time $k$, the treatment $A_k$ received at time $k$ may depend on past treatment and covariates $\bar{A}_{k-1}$ and $\bar{L}_{k}$, but that, conditional on these, it does not depend on the remaining lifetime that would apply if all future treatments were set to any particular values $\underline{a}_{k}$.

\section{Compatibility of MSMs used in the two approaches}

To obtain expressions (28) and (29) in the main text we use the result that the probability $\Pr(Y_2^{\bar A_1=a}=0|Y_1^{A_0=a}=0)$ can be written 
{\small
\begin{equation}
\begin{split}
     \Pr(Y_2^{\bar A_1=a}=0|&Y_1^{A_0=a}=0)=  \sum_{l_0}\Pr(Y_2^{\bar A_1=a}=0|Y_1^{A_0=a}=0,L_0=l_0)\frac{\Pr(Y_1=0|A_0=a,L_0=l_0)\Pr(L_0=l_0)}{\sum_{l_0^{\prime}}\Pr(Y_1=0|A_0=a,L_0=l_0^{\prime})\Pr(L_0=l_0^{\prime})}
\end{split}
\end{equation}}
and the first term in the sum can then be written
{\small
\begin{equation}
\begin{split}
     \Pr(Y_2^{\bar A_1=a}=0|&Y_1^{A_0=a}=0,L_0)
     =  \sum_{l_1}\Pr(Y_2=0|Y_1=0,\bar A_1=a,L_0,L_1=l_1)\Pr(L_1=l_1|Y_1=0,A_0=a,L_0).
\end{split}
\end{equation}}

\begin{table}[]
    \caption{Simulation results. Summary of number of individuals observed at each time, number always treated and never treated, and corresponding percentages. Numbers and percentages are the mean across the 1000 simulations.}
    \label{tab:sim.datasummary}
    \centering
    {\small
    \begin{tabular}{lllllllllllll}
    \hline
    &&\multicolumn{5}{c}{MSM-IPTW}&&\multicolumn{5}{c}{Sequential trials}\\
    &&\multicolumn{5}{c}{Time $k$}&&\multicolumn{5}{c}{Time $k$}\\
    &&0&1&2&3&4&&0&1&2&3&4\\
    \hline
\multicolumn{7}{l}{{\bf Scenario 1}}\\
Observed at time $k$&N& 1000 & 829 & 694 & 588 & 502 & & 2264 & 1414 & 873 & 509 & 238 \\ 
&\% & 100 & 83 & 69 & 59 & 50 && 100 & 62 & 39 & 22 & 10 \\ 
\addlinespace
Always treated from time 0 to $k$&N& 279 & 237 & 204 & 177 & 155 && 643 & 514 & 397 & 282 & 155 \\ 
&\%& 28 & 29 & 29 & 30 & 31 && 28 & 36 & 46 & 55 & 65 \\ 
\addlinespace
Never treated from time 0 to $k$& N & 721 & 424 & 249 & 144 & 82 & & 1621 & 900 & 475 & 227 & 82 \\ 
&\% & 72 & 51 & 36 & 25 & 16 && 72 & 64 & 54 & 45 & 35 \\ 
&&&&&&&&&&&\\
\multicolumn{7}{l}{{\bf Scenario 2}}\\
Observed at time $k$&N& 1000 & 821 & 675 & 556 & 458 && 3180 & 2178 & 1403 & 806 & 349 \\ 
&\% & 100 & 82 & 68 & 56 & 46 && 100 & 68 & 44 & 25 & 11 \\ 
\addlinespace
Always treated from time 0 to $k$&N&  53 & 45 & 39 & 34 & 29 && 197 & 142 & 98 & 61 & 29 \\ 
&\%& 5 & 5 & 6 & 6 & 6 && 6 & 7 & 7 & 8 & 8\\ 
\addlinespace
Never treated from time 0 to $k$& N & 947 & 731 & 560 & 425 & 319 && 2983 & 2036 & 1305 & 745 & 319 \\ 
&\% & 95 & 89 & 83 & 76 & 70 && 94 & 93 & 93 & 92 & 92 \\
&&&&&&&&&&&\\
\multicolumn{7}{l}{{\bf Scenario 3}}\\
Observed at time $k$&N& 1000 & 832 & 701 & 596 & 510 & & 2083 & 1333 & 865 & 543 & 287 \\ 
&\% & 100 & 83 & 70 & 60 & 51 && 100 & 64 & 42 & 26 & 14 \\ 
Always treated from time 0 to $k$&N&  387 & 326 & 278 & 241 & 210 & & 700 & 563 & 446 & 337 & 210 \\ 
&\%& 39 & 39 & 40 & 40 & 41 && 34 & 42 & 52 & 62 & 73\\ 
Never treated from time 0 to $k$& N & 613 & 352 & 213 & 129 & 77 && 1383 & 770 & 419 & 206 & 77\\ 
&\% & 61 & 42 & 30 & 22 & 15 && 66 & 58 & 48 & 38 & 27 \\ 
  \hline
    \end{tabular}}
\end{table}

\begin{figure}
    \centering
        \caption{Simulation results: bias in estimation of the risk difference using the sequential trials analysis and the MSM-IPTW analysis. The black line shows the bias at each time point and the grey area shows the Monte-Carlo 95\% CI at each time point. The MSM-IPTW results are from the MSM not conditional on $L_0$.}
    \label{fig:sim.bias}
    \includegraphics[scale=0.6]{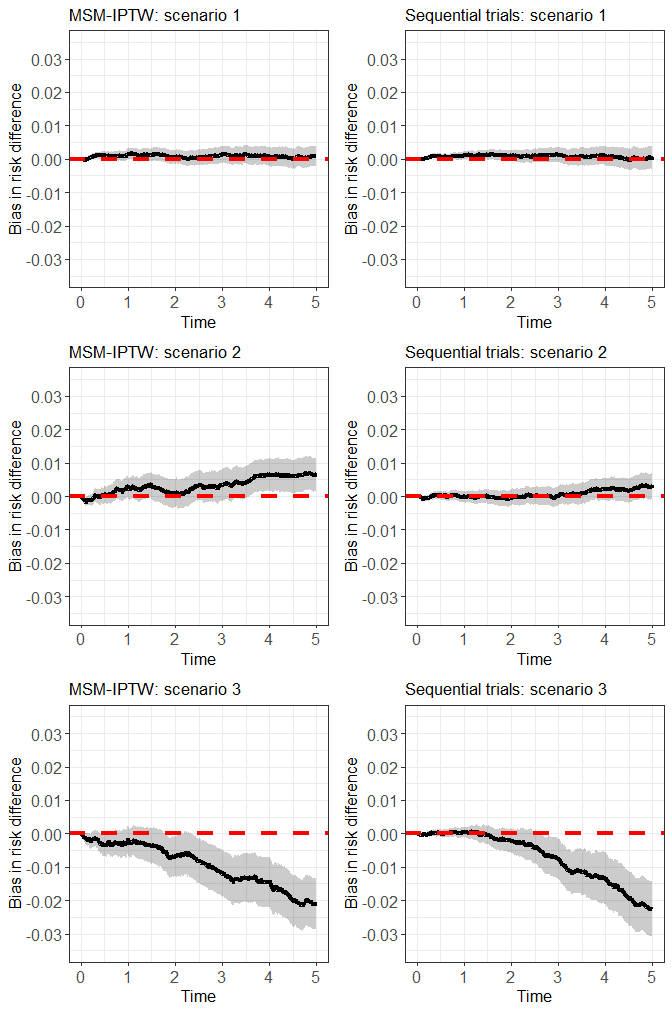}
\end{figure}

\begin{figure}
    \centering
        \caption{Simulation results: relative efficiency  of the sequential trials analysis compared with the MSM-IPTW analysis, defined as the ratio of the empirical variances of the risk difference estimates at each time. The MSM-IPTW results are from the MSM not conditional on $L_0$.}
    \label{fig:sim.eff}
    \includegraphics[scale=0.45]{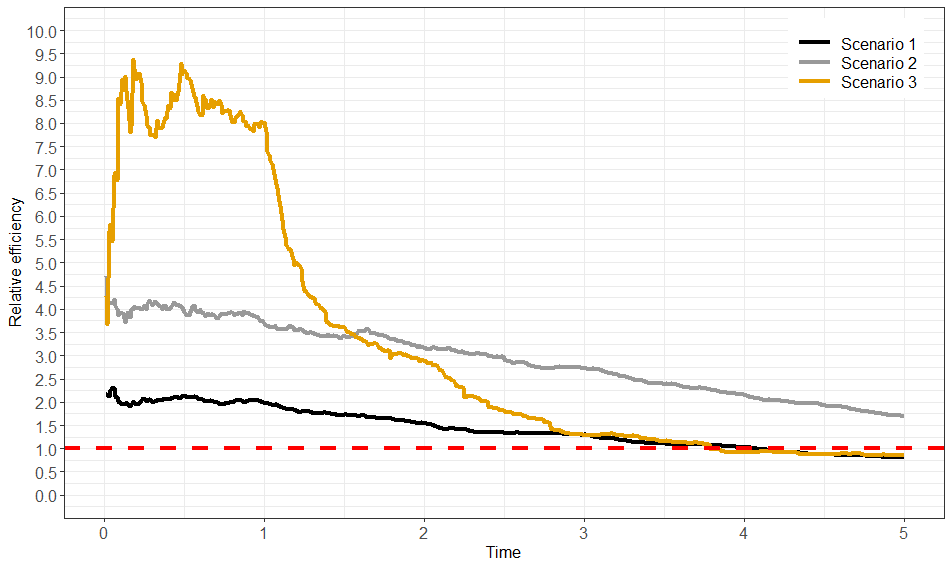}
\end{figure}

\begin{figure}
    \centering
        \caption{Simulation results using truncated weights: bias in estimation of the risk difference using the sequential trials analysis and the MSM-IPTW analysis. The black line shows the bias at each time point and the grey area shows the Monte-Carlo 95\% CI at each time point. The MSM-IPTW results are from the MSM conditional on $L_0$.}
    \label{fig:sim.bias}.
    \includegraphics[scale=0.6]{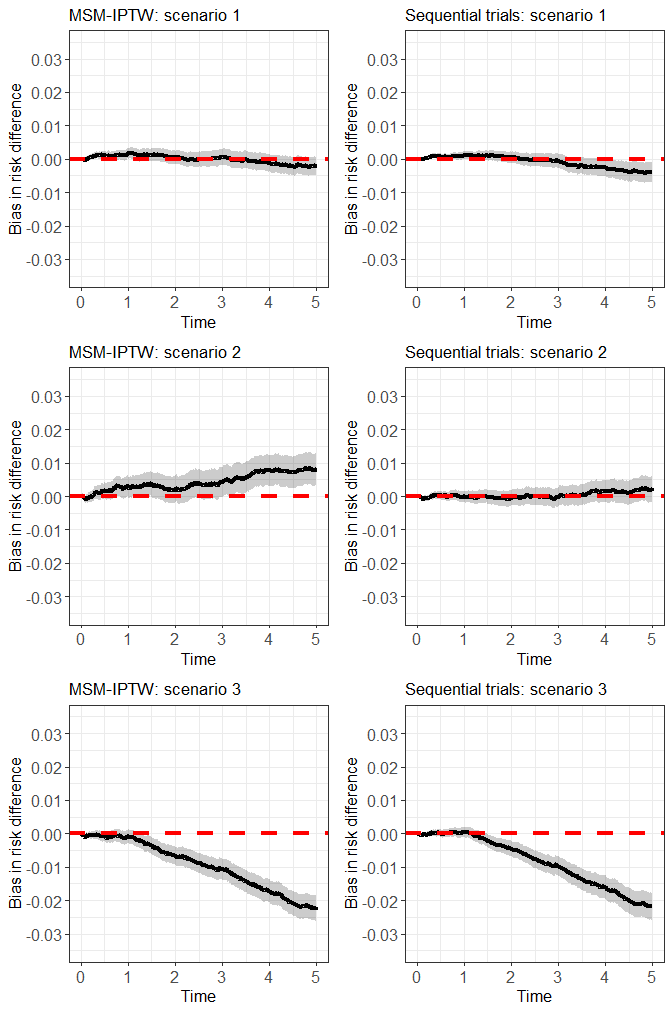}
\end{figure}

\begin{figure}
    \centering
        \caption{Simulation results using truncated weights: relative efficiency  of the sequential trials analysis compared with the MSM-IPTW analysis, defined as the ratio of the empirical variances of the risk difference estimates at each time. The MSM-IPTW results are from the MSM conditional on $L_0$.}
    \label{fig:sim.eff}
    \includegraphics[scale=0.45]{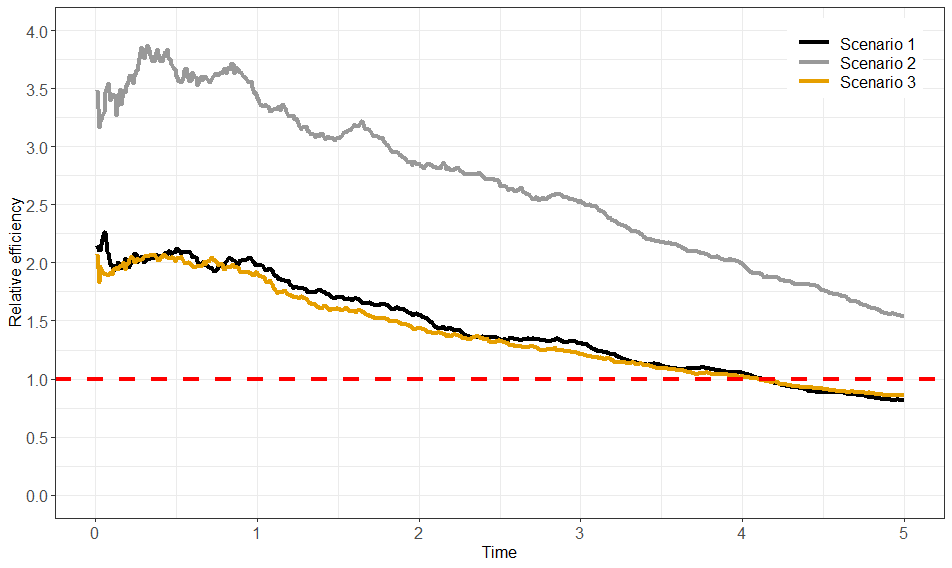}
\end{figure}

\begin{figure}
    \centering
    \includegraphics[scale=0.45]{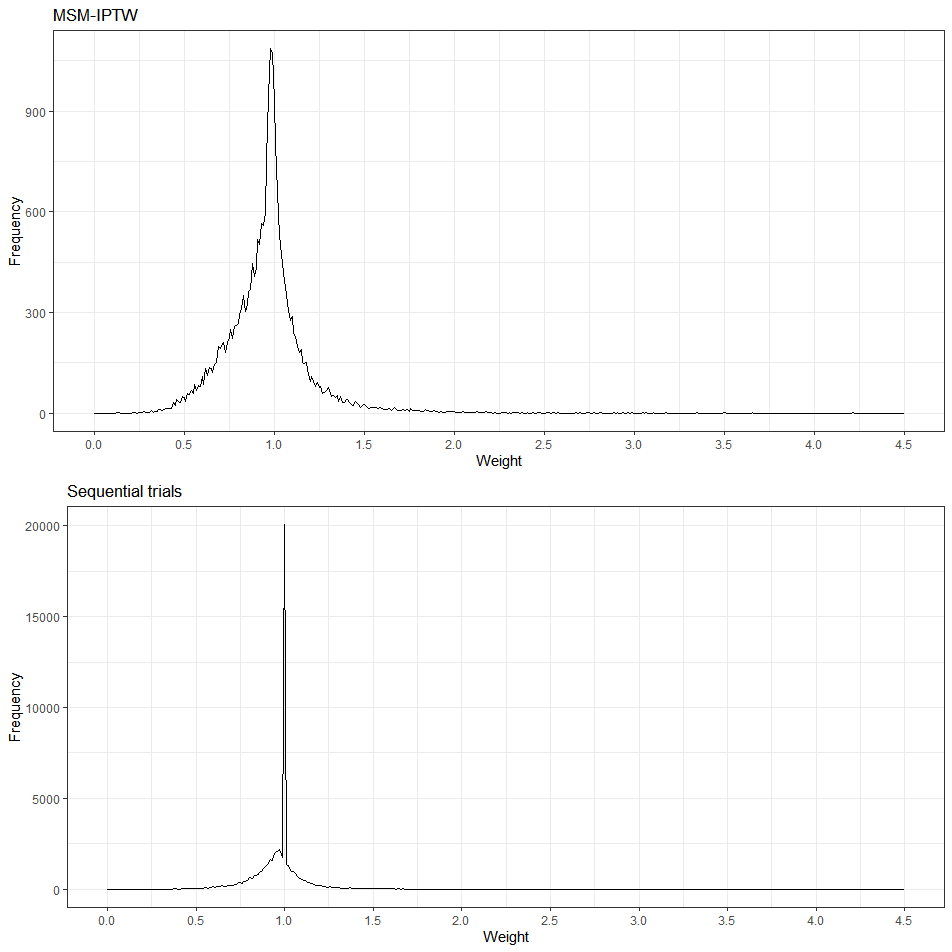}
    \caption{Plots showing the distribution of the weights used in the MSM-IPTW and sequential trials analyses of the application using UK CF Registry data.}
    \label{fig:weights.CF}
\end{figure}

\input{supp_refs.bbl}

%% file: seq_trials_arxiv_v1.bbl
\begin{thebibliography}{10}
\providecommand \doibase [0]{http://dx.doi.org/}%

\bibitem{Robins:2000}
Robins J, Hern\'{a}n M, Brumback B. Marginal structural models and causal
  inference in epidemiology.. {\it Epidemiology} 2000\string; 11\string:
  550--560.

\bibitem{Hernan:2000}
Hern\'{a}n M, Brumback B, Robins J. Marginal Structural Models to Estimate the
  Causal Effect of Zidovudine on the Survival of HIV-Positive Men. {\it
  Epidemiology} 2000\string; 11\string: 561--569.

\bibitem{Clare:2019}
Clare P, Dobbins T, Mattick R. Causal models adjusting for time-varying
  confounding -- a systematic review of the literature. {\it International
  Journal of Epidemiology} 2019\string; 48\string: 254-265.

\bibitem{Hernan:2008}
Hern\'{a}n M, Alonso A, Logan R, et al. Observational studies analyzed like
  randomized experiments: an application to postmenopausal hormone therapy and
  coronary heart disease.. {\it Epidemiology} 2008\string; 19\string: 766--779.

\bibitem{Gran:2010}
Gran J, R{\o}ysland K, Wolbers M, et al. A sequential Cox approach for
  estimating the causal effect of treatment in the presence of time-dependent
  confounding applied to data from the Swiss HIV Cohort Study. {\it Statistics
  in Medicine} 2010\string; 29\string: 2757-2768.

\bibitem{Danaei:2013SMMR}
Danaei G, Rodríguez L, Cantero O, Logan R, Hern\'{a}n M. Observational data
  for comparative effectiveness research: An emulation of randomised trials of
  statins and primary prevention of coronary heart disease. {\it Statistical
  Methods in Medical Research} 2013\string; 22\string: 70--96.

\bibitem{Clark:2015}
Clark AJ, Salo P, Lange T, et al. Onset of impaired sleep as a predictor of
  change in health-related behaviours; analysing observational data as a series
  of non-randomized pseudo-trials. {\it International journal of epidemiology}
  2015\string; 44(3)\string: 1027--1037.

\bibitem{Bhupathiraju:2017}
Bhupathiraju S, Grodstein F, Rosner B, Stampfer M, Hu F, Willett W. Hormone
  Therapy Use and Risk of Chronic Disease in the Nurses Health Study: A
  Comparative Analysis With the Women's Health Initiative. {\it American
  Journal of Epidemiology} 2017\string; 186\string: 696--708.

\bibitem{Suttorp:2015}
Suttorp MM, Hoekstra T, Mittelman M, et al. Treatment with high dose of
  erythropoiesis-stimulating agents and mortality: Analysis with a sequential
  Cox approach and a marginal structural model.. {\it Pharmacoepidemiology and
  Drug Safety} 2015\string; 24\string: 1068-1075.

\bibitem{Thomas:2020}
Thomas LE, Yang S, Wojdyla D, Schaubel DE. Matching with time-dependent
  treatments: A review and look forward. {\it Statistics in Medicine} 2020.

\bibitem{Hernan:2016AJE}
Hern\'{a}n M, Robins J. Using Big Data to Emulate a Target Trial When a
  Randomized Trial Is Not Available. {\it American Journal of Epidemiology}
  2016\string; 183\string: 758--764.

\bibitem{Hernan:2016JCE}
Hern\'{a}n M, Sauer B, Hernandez-Diaz S, Platt R, Shrier I. Specifying a target
  trial prevents immortal time bias and other self-inflicted injuries in
  observational analyses. {\it Journal of Clinical Epidemiology} 2016\string;
  79\string: 70--75.

\bibitem{Garcia:2017}
Garcia-Albeniz X, Hsu J, Hern\'{a}n M. The value of explicitly emulating a
  target trial when using real world evidence: an application to colorectal
  cancer screening. {\it European Journal of Epidemiology} 2017\string;
  32\string: 495-500.

\bibitem{Hernan:2018}
Hern\'{a}n M. How to estimate the effect of treatment duration on survival
  outcomes using observational data.. {\it British Medical Journal}
  2018\string; 360\string: k182.

\bibitem{Didelez:2016}
Didelez V. Commentary: Should the analysis of observational data always be
  preceded by specifying a target experimental trial?. {\it International
  journal of epidemiology} 2016\string; 45(6)\string: 2049--2051.

\bibitem{Petersen:2014}
Petersen ML, {van der Laan} MJ. Causal models and learning from data:
  integrating causal modeling and statistical estimation. {\it Epidemiology
  (Cambridge, Mass.)} 2014\string; 25(3)\string: 418.

\bibitem{Gran:2019}
Gran J, Aalen O. Letter to the Editor: Comparison of statistical approaches
  dealing with time-dependent confounding in drug effectiveness studies (SMMR,
  Vol. 27, Issue 6, 2018). {\it Statistical Methods in Medical Research}
  2019\string; 28\string: 321-322.

\bibitem{Karim:2018}
Karim M, Petkau J, Gustafson P, et al. Comparison of statistical approaches
  dealing with time-dependent confounding in drug effectiveness studies. {\it
  Statistical Methods in Medical Research} 2018\string; 27\string: 1709--1722.

\bibitem{Martinussen:2013}
Martinussen T, Vansteelandt S. On collapsibility and confounding bias in Cox
  and Aalen regression models. {\it Lifetime data analysis} 2013\string;
  19(3)\string: 279--296.

\bibitem{Hernan:2010}
Hern\'{a}n M. The Hazards of Hazard Ratios. {\it Epidemiology} 2010\string;
  21\string: 13--15.

\bibitem{Aalen:2015}
Aalen OO, Cook RJ, R{\o}ysland K. Does Cox analysis of a randomized survival
  study yield a causal treatment effect?. {\it Lifetime data analysis}
  2015\string; 21(4)\string: 579--593.

\bibitem{Martinussen:2020}
Martinussen T, Vansteelandt S, Andersen P. Subtleties in the interpretation of
  hazard contrasts. {\it Lifetime Data Analysis} 2020\string; 36\string:
  833--855.

\bibitem{Keogh:2021}
Keogh R, Seaman S, Gran J, Vansteelandt S. Simulating longitudinal data from
  marginal structural models using the additive hazard model. {\it Biometrical
  Journal} 2021.

\bibitem{Petersen:2007a}
Petersen M, Deeks S, Martin J, {van der Laan} M. History-adjusted Marginal
  Structural Models for Estimating Time-varying Effect Modification. {\it
  American Journal of Epidemiology} 2007\string; 166\string: 985--993.

\bibitem{Robins:2007}
Robins J, Hern\'{a}n MA, Rotnitzky A. Invited Commentary: Effect Modification
  by Time-varying Covariates. {\it American Journal of Epidemiology}
  2007\string; 166\string: 994-1002.

\bibitem{Petersen:2007b}
Petersen M, {van der Laan} M. {Petersen et al. Respond to "Effect modification
  by time-varying covariates"}. {\it American Journal of Epidemiology}
  2007\string; 166\string: 1003-1004.

\bibitem{CFT:2019}
{Cystic Fibrosis Trust} . {UK Cystic Fibrosis Registry Annual Data Report
  2019}.  2020.

\bibitem{Rowe:2005}
Rowe S, Miller S, Sorscher E. Cystic Fibrosis. {\it New England Journal of
  Medicine} 2005\string; 352\string: 1992--2001.

\bibitem{MacNeill:2016}
MacNeill SJ. {\it Hodson and Geddes' Cystic Fibrosis}ch.~Epidemiology of Cystic
  Fibrosis\string: 18--40; CRC Press .
\newblock 2016.

\bibitem{Dodge:2007}
Dodge J, Lewis P, Stanton M, Wilsher J. Cystic fibrosis mortality and survival
  in the UK: 1947-2003.. {\it European Respiratory Journal} 2007\string;
  29\string: 522--526.

\bibitem{Keogh:2018JCF2}
Keogh R, Szczesniak R, Taylor-Robinson D, Bilton D. Up-to-date and projected
  estimates of survival for people with cystic fibrosis using baseline
  characteristics: A longitudinal study using UK patient registry data.. {\it
  Journal of Cystic Fibrosis} 2018\string; 17\string: 218--227.

\bibitem{Pressler:2008}
Pressler T. Review of recombinant human deoxyribonuclease (rhDNase) in the
  management of patients with cystic fibrosis. {\it Biologics} 2008\string;
  2\string: 611--617.

\bibitem{Yang:2018}
Yang C, Montgomery M. Dornase alfa for cystic fibrosis. {\it Cochrane Database
  of Systematic Reviews} 2018(9).

\bibitem{Newsome:2018SIM}
Newsome S, Keogh R, Daniel R. Estimating long-term treatment effects in
  observational data: A comparison of the performance of different methods
  under real-world uncertainty. {\it Statistics in Medicine} 2018\string;
  37\string: 2367--2390.

\bibitem{Newsome:2019JCF}
Newsome S, Daniel R, Carr S, Bilton D, Keogh R. Investigating the effects of
  long-term dornase alfa use on lung function using registry data. {\it Journal
  of Cystic Fibrosis} 2019\string; 110--117.

\bibitem{Taylor-Robinson:2017}
Taylor-Robinson D, Archangelidi O, Carr S, et al. Data Resource Profile: The UK
  Cystic Fibrosis Registry.. {\it International Journal of Epidemiology}
  2017\string; 47\string: 9--10e.

\bibitem{Royston:2013}
Royston P, Parmar M. Restricted mean survival time: an alternative to the
  hazard ratio for the design and analysis of randomized trials with a
  time-to-event outcome. {\it BMC Medical Research Methodology} 2013\string;
  152\string: 152.

\bibitem{Cox:1972}
Cox D. Regression Models and Life-Tables. {\it Journal of the Royal Statistical
  Society (Series B)} 1972\string; 34\string: 187--220.

\bibitem{Aalen:1989}
Aalen O. {A linear regression model for the analysis of life times}. {\it
  Statistics in Medicine} 1989\string; 8\string: 907--925.

\bibitem{Aalen:2008}
Aalen O, Borgan {\O}, Gjessing H. {\it Survival and Event History Analysis: A
  Process Point of View.}
\newblock New York: Springer .
\newblock 2008.

\bibitem{Vanderweele:2009}
VanderWeele T. Concerning the consistency assumption in causal inference. {\it
  Epidemiology} 2009\string; 20\string: 880-883.

\bibitem{Daniel:2013}
Daniel R, Cousens S, {De Stavola} B, Kenward M, Sterne J. Methods for dealing
  with time-dependent confounding. {\it Statistics in Medicine} 2013\string;
  32\string: 1584--1618.

\bibitem{D'Agostino:1990}
D'Agostino R, Lee ML, Belanger A, Cupples L, Anderson K, Kannel W. Relation of
  pooled logistic regression to time-dependent Cox regression analysis: The
  Framingham Heart Study. {\it Statistics in Medicine} 1990\string; 9\string:
  1501--1515.

\bibitem{Roysland:2011}
R{\o}ysland K, Gran J, Ledergerber B, Wyl vV, Young J, Aalen O. Analyzing
  direct and indirect effects of treatment using dynamic path analysis applied
  to data from the Swiss HIV Cohort. {\it Statistics in Medicine} 2011\string;
  30\string: 2947–2958.

\bibitem{Richardson:2014}
Richardson T, Rotnitzky A. Causal Etiology of the Research of James M. Robins.
  {\it Statisticl Science} 2014\string; 29\string: 459-484.

\bibitem{HernanRobins:2010}
Hern\'{a}n M, Robins J. {\it Causal Inference: What If.}
\newblock Boca Raton: Chapman and Hall/CRC .
\newblock 2010.

\bibitem{Morris:2019}
Morris T, White I, Crowther M. Using simulation studies to evaluate statistical
  methods. {\it Statistics in Medicine} 2019\string; 38\string: 2074--2102.

\bibitem{McKone:2006}
McKone E, Goss C, Aitken M. CFTR Genotype as a Predictor of Prognosis in Cystic
  Fibrosis. {\it Chest} 2006\string; 130\string: 1141--1147.

\bibitem{Cole:2017}
Cole S, Edwards J, Hall H, et al. Incident AIDS or Death After Initiation of
  Human Immunodeficiency Virus Treatment Regimens Including Raltegravir or
  Efavirenz Among Adults in the United States. {\it Clinical Infectious
  Diseases} 2017\string; 64\string: 1591--1596.

\bibitem{vanderLaan:2005}
{van der Laan} M, Petersen M, Joffe M. History-adjusted marginal structural
  models and statically-optimal dynamic treatment regimens.. {\it International
  Journal of Biostatistics} 2005\string; 1\string: article 4.

\bibitem{Bembom:2009}
Bembom O, {van der Laan} M, Haight T, Tager I. Leisure-time Physical Activity
  and All-cause Mortality in an Elderly Cohort. {\it Epidemiology} 2009\string;
  20\string: 424–430.

\bibitem{Schaubel:2006}
Schaubel D, Wolfe R, Port F. A sequential stratification method for estimating
  the effect of a time-dependent experimental treatment in observational
  studies.. {\it Biometrics} 2006\string; 62\string: 910–917.

\bibitem{Seaman:2020}
Seaman S, Dukes R, Vansteelandt S. Adjusting for time‐varying confounders in
  survival analysis using structural nested cumulative survival time models.
  {\it Biometrics} 2019.

\bibitem{Sawicki:2012}
Sawicki G, Signorovitch J, Zhang J, Latremouille-Viau M, Wu E, Shi L. Reduced
  mortality in cystic fibrosis patients treated with tobramycin inhalation
  solution. {\it Pediatric Pulmonology} 2012\string; 47\string: 44--52.

\bibitem{Cole:2008}
Cole S, Hern\'{a}n M. Constructing inverse probability weights for marginal
  structural models. {\it American Journal of Epidemiology} 2008\string;
  168\string: 656--664.

\end{thebibliography}


\begin{thebibliography}{}

\bibitem[\protect\citeauthoryear{Cole and Hernan}{Cole and
  Hernan}{2008}]{Cole:2008}
Cole, S. and Hernan, M. (2008).
\newblock Constructing inverse probability weights for marginal structural
  models.
\newblock {\em American Journal of Epidemiology} {\bf 168,} 656--664.

\bibitem[\protect\citeauthoryear{Daniel, Cousens, De~Stavola, Kenward, and
  Sterne}{Daniel et~al.}{2013}]{Daniel:2013}
Daniel, R., Cousens, S., De~Stavola, B., Kenward, M., and Sterne, J. (2013).
\newblock Methods for dealing with time-dependent confounding.
\newblock {\em Statistics in Medicine} {\bf 32,} 1584--1618.

\end{thebibliography}
